\documentclass[preprint,12pt,authoryear]{elsarticle}

\usepackage{amssymb}
\usepackage{amsmath}
\usepackage{booktabs}
\usepackage{multirow}
\usepackage{tabularx} 
\usepackage{graphicx}
\usepackage{rotating}  
\usepackage{placeins}
\usepackage{afterpage}
\usepackage{caption} 
\begin{document}

\begin{frontmatter}



%\title{FFV-PINN: A Fast Physics-Informed Neural Network Incorporating Simplified Finite Volume Method and Residual Correction Loss}
\title{FFV-PINN: A Fast Physics-Informed Neural Network with Simplified Finite Volume Discretization and Residual Correction}


\author[inst1]{Chang Wei}
\author[inst1]{Yuchen Fan}
\author[inst2]{Jian Cheng Wong}
\author[inst2]{Chin Chun Ooi}
\author[inst1]{Heyang Wang\corref{bbb}}
\author[inst2]{Pao-Hsiung Chiu\corref{aaa}}

\cortext[bbb]{Corresponding author at: School of Mechanical Engineering, Tianjin University, Tianjin, 300350, China. \textit{E-mail address:} heyang.wang@tju.edu.cn (H. Wang).}
\cortext[aaa]{Corresponding author at: Institute of High Performance Computing, Agency for Science, Technology and Research (A*STAR), 138632, Singapore. \textit{E-mail address:} chiuph@a-star.edu.sg (P-H. Chiu).}

\affiliation[inst1]{organization={School of Mechanical Engineering, Tianjin University},%Department and Organization
            % addressline={Address One}, 
            city={Tianjin},
            postcode={300350}, 
            % state={State One},
            country={China}}
            
\affiliation[inst2]{organization={Institute of High Performance Computing, Agency for Science, Technology and Research (A*STAR)},%Department and Organization
            postcode={138632}, 
            country={Singapore}}

\begin{abstract}

With the growing application of deep learning techniques in computational physics, physics-informed neural networks (PINNs) have emerged as a major research focus. However, today's PINNs encounter several limitations. Firstly, during the construction of the loss function using automatic differentiation, PINNs often neglect information from neighboring points, which hinders their ability to enforce physical constraints and diminishes their accuracy. Furthermore, issues such as instability and poor convergence persist during PINN training, limiting their applicability to complex fluid dynamics problems. To address these challenges, this paper proposes a fast physics-informed neural network framework that integrates a simplified finite volume method (FVM) and residual correction loss term, referred to as Fast Finite Volume PINN (FFV-PINN). FFV-PINN utilizes a simplified FVM discretization for the convection term, which is one of the main sources of instability, with an accompanying improvement in the dispersion and dissipation behavior. Unlike traditional FVM, which requires careful selection of an appropriate discretization scheme based on the specific physics of the problem such as the sign of the convection term and relative magnitudes of convection and diffusion, the FFV-PINN outputs can be simply and directly harnessed to approximate values on control surfaces, thereby simplifying the discretization process.
% without sacrificing accuracy with improvement of dispersion and dissipation behavior.
Moreover, a residual correction loss term is introduced in this study that significantly accelerates convergence and improves training efficiency. To validate the performance of FFV-PINN, we solve a series of challenging problems --- including flow in the two-dimensional steady and unsteady lid-driven cavity, three-dimensional steady lid-driven cavity, backward-facing step scenarios, and natural convection at previously unsurpassed Reynolds ($Re$) number and Rayleigh ($Ra$) number, respectively --- that are typically difficult for PINNs. Notably, the FFV-PINN can achieve data-free solutions for the lid-driven cavity flow at $Re = 10000$ and natural convection at $Ra = 10^8$ for the first time in PINN literature, even while requiring only 680s and 231s respectively. These results further highlight the effectiveness of FFV-PINN in improving both speed and accuracy, marking another step forward in the progression of PINNs as competitive neural PDE solvers. 
% Notably, we achieve data-free solutions for the lid-driven cavity flow at $Re = 10000$ for the first time in just 10 minutes, demonstrating the efficiency of FFV-PINN.

\end{abstract}



\end{frontmatter}


\section{Introduction}
Over the past few years, data-driven models have increasingly demonstrated their utility across various fields \citep{lecun2015deep}, including but not limited to computer vision, natural language processing, recommendation systems, and autonomous driving. However, when encountering complex physical systems, models that rely exclusively on data often struggle to accurately capture the underlying governing patterns \citep{karniadakis2021physics}. In this context, physics-informed neural networks (PINNs) \citep{RAISSI2019686} have emerged as a highly promising and innovative solution. The integration of the representational capacity of neural networks with prior physical knowledge enables PINNs to mitigate a fundamental drawback of purely data-driven models, namely, their reliance on extensive and often impractical datasets. Excitingly, this synergy promises to enhance the model's capability to solve complex problems with fewer data points, making it particularly suitable for real-world applications where data can be scarce or difficult to obtain. In addition to these advantages, PINNs exhibit remarkable versatility, making them applicable to both forward and inverse problems \citep{yuan2022pinn,lou2021physics}. The ability to handle both problem types within a unified framework further extends their applicability across diverse fields including fluid dynamics \citep{cai2021physics}, structural mechanics \citep{hu2024physics}, climate modeling \citep{dutta2024aq}, and biomedical simulations \citep{sarabian2022physics}.

% Forward problems involve predicting a system's behavior based on known initial and boundary conditions. In these scenarios, PINNs utilize the governing equations of the system to simulate its evolution over time or in response to specific inputs. This capability is essential for comprehending the dynamics of complex systems and forecasting their future states.  Conversely, inverse problems present greater challenges than forward modeling, as they require inferring missing system information from limited or even noisy data. PINNs, in contrast to traditional numerical methods, demonstrate superiority in this domain by integrating data-driven approaches with physical constraints for accurate parameter estimation and data reconstruction. Consequently, the ability of PINNs to address both forward and inverse problems within a unified framework not only simplifies the modeling paradigm but also broadens their applicability, spurring widespread adoption across a multitude of disciplines, including fluid dynamics \citep{cai2021physics}, structural mechanics \citep{hu2024physics}, climate modeling \citep{dutta2024aq}, and biomedical simulations \citep{sarabian2022physics}.

However, despite their remarkable potential across a wide range of physical problems, PINNs encounter several challenges that impede their broader development and practical implementation  \citep{wang2022and}. A primary limitation stems from their relatively weak enforcement of physical constraints  \citep{chiu2022can,cvpinn}. Specifically, the loss function in PINNs is typically formulated using automatic differentiation (AD) by solely computing the residuals of the governing equations at isolated collocation points \citep{RAISSI2019686}. While AD is simple and straightforward to implement, it does not incorporate information from adjacent points, thereby neglecting the conservation relationships that exist among them \citep{karnakov2024solving}. 
% In numerous physical systems, conservation relationships are essential for accurately representing the behavior of the investigated model system. For instance, in fluid dynamics, conservation laws such as mass conservation and momentum conservation dictate that the state at one point must satisfy certain constraints in relation to adjacent points \citep{darwish2016finite}.
When PINNs fail to account for these relationships, the resulting solutions may violate fundamental physical principles \citep{jagtap2020conservative}, particularly in complex scenarios where interactions among points play a critical role. Although increasing the number of collocation points may improve the model's adherence to physical laws and enhance solution accuracy, this strategy incurs significantly higher computational costs and longer training times. 

Several strategies have been proposed to enhance the conservation properties of PINNs. For instance, the conservative PINN (cPINN) \citep{jagtap2020conservative} employs domain decomposition with interfacial flux conservation; however, this approach may suffer from interface errors that impair training stability and solution accuracy. Alternatively, two-stage PINNs \citep{lin2022two} attempt to improve conservation by incorporating the mean square error of conserved quantities into the loss function, and this idea has been further refined into a single-stage formulation using adaptive weighting techniques \citep{wang2022modified}. The PINN-Proj model \citep{baez2024guaranteeing} enforces conservation by projecting the network output onto a conservation-adhering function space, albeit limited to systems with zero net flux. Other approaches include Continuity Planes and Monte Carlo integration \citep{gurieva2022application}, tailored loss functions for solitons \citep{wu2022prediction}, and direct integration of integral conservation laws via finite volume methods \citep{hansen2023learning}. Integral formulations have also been explored, either by predicting integral quantities directly \citep{rajvanshi2024integral} or by embedding weak-form representations into the loss via coupled neural networks \citep{wang2024coupled}. Nevertheless, many of these methods rely on initial conditions for calculating conserved quantities, rendering them unsuitable for steady problems. Furthermore, the complex derivation and computation of conserved quantities often restrict their applicability \citep{lin2022two}. Critically, these approaches are predominantly tailored for integrable systems and may not be directly applicable to complex equations such as the Navier-Stokes equations.

Beyond the aforementioned issue, PINNs frequently encounter challenges pertaining to instability and poor convergence during training \citep{fabiani2024stability,cuomo2022scientific}. The optimization landscape for PINNs is highly non-convex, complicating the training process and inducing oscillations in the loss function. Such instability can severely impede the model's ability to achieve convergence towards a stable solution, especially in complex fluid dynamics problems characterized by highly nonlinear dynamics. Consequently, more efficient strategies must be developed to strengthen physical constraints while ensuring computational stability and convergence.

Recently, increasing researchers \citep{lim2022physics,xiang2022hybrid,chiu2022can,jiang2023applications,xiao2024least,cen2024deep,mei2024unified,su2024finite,zhang2025physics,zhu2024finite} have explored the integration of traditional numerical methods, such as the finite difference method (FDM) \citep{thomas2013numerical} and the finite volume method (FVM) \citep{patankar2018numerical}, within the PINN framework to mitigate the issue of weak physical constraint enforcement. These approaches leverage well-established numerical methods \citep{malalasekera2007introduction} to approximate the derivative terms in the partial differential equations (PDEs) using discretization schemes, thereby enhancing the spatial relationship across collocation points. For example, Xiang et al. \citep{xiang2022hybrid}  proposed a hybrid finite difference physics-informed neural network (HFD-PINN) framework, wherein FDM was applied locally to replace AD within the PINN framework, thereby effectively balancing computational efficiency and accuracy. Zhang et al. \citep{zhang2025physics} further proposed a physics-informed graph neural network based on the finite volume method (FVGP-Net) for the unsupervised training and prediction of steady incompressible laminar convective heat transfer problems. 
% FVGP-Net substitutes AD with FVM-based numerical differentiation, improving both training efficiency and prediction accuracy.
Chiu et al. \citep{chiu2022can} also proposed to couple AD with a numerical scheme for loss construction, resulting in fast and sample-efficient PINNs. 
% These efforts suggest that integrating established numerical methods into the PINN framework offers a promising avenue for addressing the challenge of weak physical constraint enforcement, which is a prevalent issue in the current training of PINNs.

Concurrently, researchers have proposed various methods and strategies, such as adaptive training techniques \citep{katende2024some,fabiani2024stability,meng2024physics,jahani2024enhancing,rathore2024challenges}, modified loss function formulations \citep{meng2023pinn,hu2023solution,nosrati2024manipulating}, as well as advanced neural network architectures \citep{kaplarevic2023identifying,wang2024pinn,moser2023modeling}, to address inherent challenges of stability and convergence when training PINNs in the AD framework. For instance, McClenny et al. \citep{mcclenny2023self} improved model stability by optimizing the physical residual weights in the loss function through adaptive weighting strategies.

Despite these advancements, challenges remain in ensuring the conservation constraints are adequately satisfied throughout the training process and in maintaining the model's stability and convergence. Firstly, existing hybrid frameworks, which integrate PINNs with traditional numerical methods, frequently encounter numerical instability when discretizing convection terms. This instability manifests as pronounced dispersion and dissipation errors, which not only compromise the accuracy of the solutions but also substantially obstruct the effective enforcement of physical constraints. As a result, these models struggle to accurately capture complex physical phenomena \citep{wang2023solution,jiang2023applications}. Secondly, prevailing strategies for enhancing the convergence and stability of PINNs predominantly adopt machine learning methods, emphasizing algorithmic or architectural improvements while giving insufficient attention to the physical constraints. This limited incorporation of physical information into the optimization process restricts the model’s stability and convergence, particularly in complex physical scenarios.

\begin{figure}[!ht]
    \centering
    \includegraphics[width=1\linewidth]{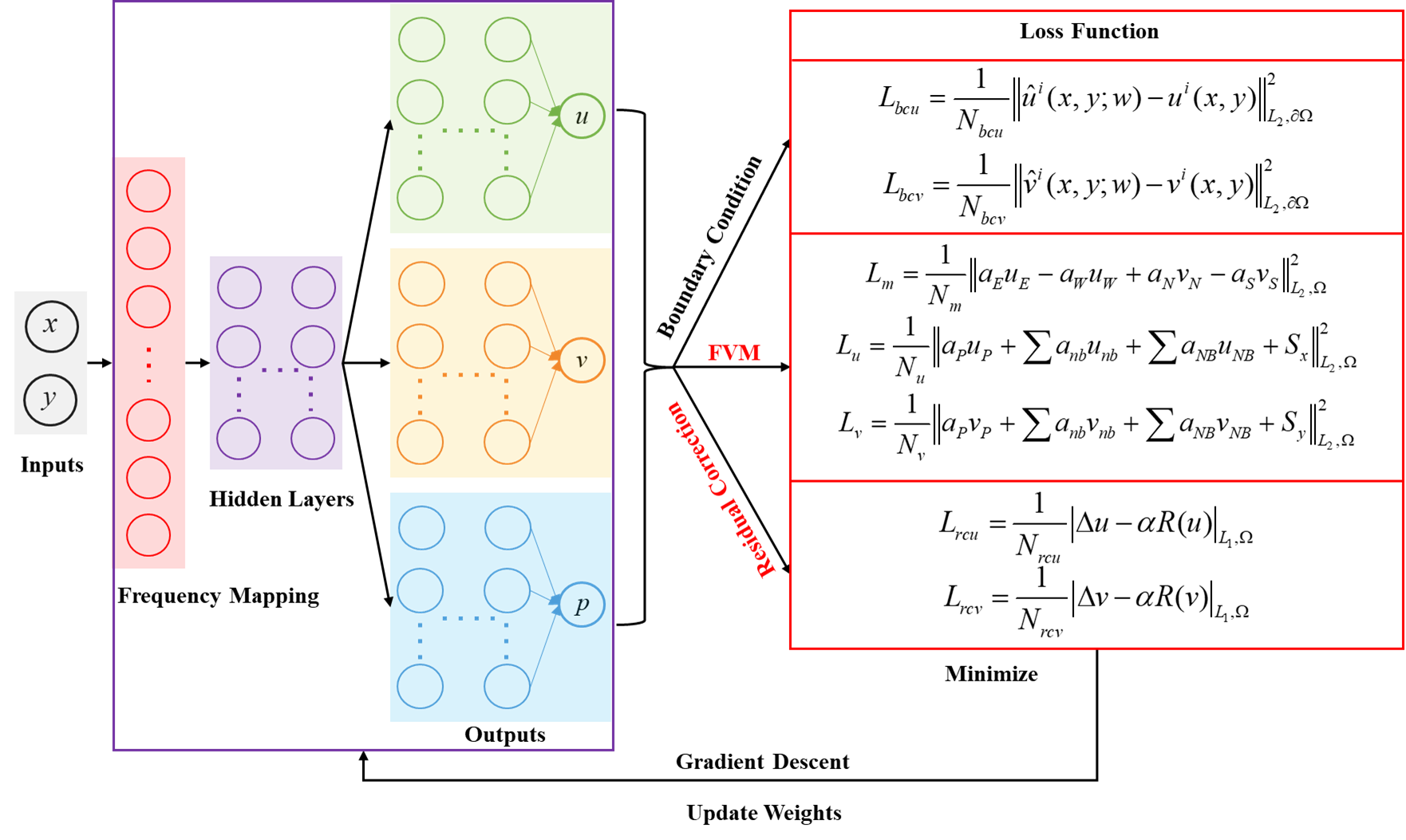}
    \caption{Schematic representation of the proposed FFV-PINN framework integrated with the simplified FVM and residual correction loss. The neural network takes spatial coordinates $(x,y)$ as inputs and predicts velocity components $u,v$ and pressure $p$. The loss function comprises three components: the BC loss, the physical constraint loss derived from the discretized governing equations, and the residual correction loss, which refines the solution by addressing deviations from the expected physical constraints.}
    \label{fig:nn}
\end{figure}

To address the aforementioned limitations, this paper introduces a fast finite volume physics-informed neural network framework, referred to as FFV-PINN (schematic in Fig. \ref{fig:nn}), which integrates a simplified FVM to enhance the imposition of physical constraints and a novel residual correction loss to enable fast and stable training. Specifically, the simplified FVM is proposed to improve numerical stability, thereby facilitating the better enforcement of physical constraints. Distinct from traditional FVM, where discretization schemes for convective terms are selected based on the direction of the convection, as well as the relative magnitudes of convection and diffusion \citep{tao2021my}, FFV-PINN bypasses this tedious step by utilizing neural network outputs to directly approximate values on the control surfaces. In this way, the FFV-PINN can improve the dispersion and dissipation behavior while significantly simplifying the discretization process. In the meantime, the novel residual correction loss is designed from the perspective of physical constraints to enhance the convergence and stability of the neural network during training. By acting as a guiding mechanism for parameter updates, this loss term ensures that the network’s outputs remain closely aligned with established conservation principles. To evaluate FFV-PINN's performance, a series of canonical fluid dynamics benchmark problems, such as two-dimensional (2D) steady and unsteady lid-driven cavity flows, three-dimensional (3D) steady lid-driven cavity flow, backward-facing step flow, and natural convection phenomena, are investigated. These selected cases are commonly employed in the evaluation of new computational physics algorithms for their respective inherent complexity and associated challenges for numerical methods and serve as rigorous validation of the framework's predictive capabilities in this work.

Compared to previous studies that enforce conservation laws, the FFV-PINN presents notable advantages. First, by integrating the FVM, which is inherently designed for conservation laws \citep{hansen2023learning}, the discretized PDE is directly embedded into the loss function. This allows FFV-PINN to be applied to more intricate systems beyond the limitations of integrable models. Second, the residual correction loss term in FFV-PINN, derived from the conservation equation, offers more effective guidance for neural network parameter updates during training. Furthermore, unlike many existing methods, FFV-PINN does not rely on initial conditions to compute conserved quantities, making it applicable to both steady and unsteady problems.

% to enable fast and accurate PINN training.
%Specifically, in FFV-PINN, a simplified finite volume method is employed to handle the discretization of the convection term, which is one of the main sources of instability \citep{tao2021my}. 
% significantly simplifies the discretization process while maintaining accuracy.
% In the meantime, the newly introduced residual correction loss term serves as a guiding mechanism for the updates of the neural network parameters, thereby enhancing the fulfillment of physical constraints by aligning neural network outputs with established conservation principles. 
%as result, this newly introduced term serves as a guiding mechanism for the updates of the neural network parameters, thereby enhancing the stability of the model and facilitating a more efficient convergence behavior during the training process.

Our main contributions can be summarized as:
\begin{itemize}
    \item We propose a simplified FVM to handle the discretization of the convection term within governing equations, resulting in a more streamlined discretization process while maintaining numerical accuracy.
    \item We introduce a residual correction loss term designed to align the neural network outputs with conservation principles. This innovative loss term not only enforces the physical constraints more effectively but also acts as a guide for updating the neural network parameters, thereby improving the performance of the model and facilitating faster convergence during training.
    \item The efficacy of FFV-PINN is demonstrated through application to challenging benchmark problems. Notably, data-free solutions for a lid-driven cavity flow at high Reynolds number ($Re$ = 10000) and a natural convection problem at high Rayleigh number ($Ra$ = $10^8$) are presented for the first time, demonstrating high accuracy and reduced training time that other PINNs have heretofore been unable to attain.    
\end{itemize}

The remainder of this paper is structured as follows: Section \ref{sec:method} outlines the basic loss formulation of PINNs, proceeds to discuss their integration with numerical discretization techniques, and finally presents the FFV-PINN framework. Section \ref{numerical experiments} reports extensive experimental studies across multiple benchmark problems, with explicit comparison to existing literature to demonstrate FFV-PINN's advantages. Section \ref{sec:remarks} offers concluding remarks and outlines potential directions for future research.

\section{Methodology}\label{sec:method}
\subsection{The overview of PINNs} 
Without loss of generality, we consider a canonical physics-informed learning problem governed by the following general PDEs:
\begin{subequations} \label{eq:pde_ibc_eqn}
    \begin{align}
        & \text{PDE:} & \mathcal{N}_\vartheta[u(x,t)] &= h(x,t), & x\in\Omega, t\in(0,T] \label{eq:pde_eqn} \\
        & \text{IC:} & u(x,t=0) &= u_0(x), & x\in\Omega \label{eq:ic_eqn} \\
        & \text{BC:} & u(x,t) &= g(x,t), & x\in\partial\Omega, t\in(0,T] \label{eq:bc_eqn}
    \end{align}
\end{subequations} 
where the input variables are denoted as $x$ and $t$, representing the spatial and temporal variables respectively, while $u$ represents the output of interest. The general differential operator $\mathcal{N}_\vartheta[u(x,t)]$ contains parameters $\vartheta$ that are associated with the PDEs and may include both linear and nonlinear combinations of the temporal and spatial derivatives of the output $u$. In addition, $h(x,t)$ represents a general source term defined within the domain $x\in\Omega, t\in(0, T]$. The initial condition (IC), as given by Eq.~\ref{eq:ic_eqn}, specifies $u(x,t) = u_0(x)$ at the initial time $t = 0$. Furthermore, Eq.~\ref{eq:bc_eqn} specifies the boundary condition (BC), indicating that $u(x,t)$ is equal to $g(x,t)$ at the boundary of the domain $\partial\Omega$. Here, $g(x,t)$ may be a Dirichlet BC, a Neumann BC, a Robin BC, or any combination thereof.

The loss function of a PINN model is composed of multiple terms, including PDE loss, IC loss, and BC loss.
% , while in inverse problems, an additional data loss term is also incorporated:
\begin{subequations} \label{eq:physics_loss_fn}
    \begin{align}
   & \mathcal{L} = \lambda_{pde}\ \mathcal{L}_{pde} + \lambda_{ic}\ \mathcal{L}_{ic} + \lambda_{bc}\ \mathcal{L}_{bc}   \\
    &\mathcal{L}_{pde} = \frac{1}{N_{pde}}\lVert \mathcal{N}_\vartheta[f(\cdot;\boldsymbol{w})] - h(\cdot) \rVert _{L_2(\Omega \times (0,T])}^2 \label{PDE_loss}\\
    &\mathcal{L}_{ic} = \frac{1}{N_{ic}}\lVert f(\cdot,t=0;\boldsymbol{w}) - u_0(\cdot) \rVert _{L_2(\Omega)}^2 \\
    &\mathcal{L}_{bc} = \frac{1}{N_{bc}}\lVert \mathcal{B}[f(\cdot;\boldsymbol{w})] - g(\cdot) \rVert _{L_2(\partial\Omega \times (0,T])}^2 
    % &\mathcal{L}_{data} = \frac{1}{N_{data}}\lVert f(x,t;\boldsymbol{w})  - u(x,t) \rVert _{L_2(\Omega \times (0,T])}^2
    \end{align}
\end{subequations} 
where $f(\cdot;\boldsymbol{w})$ represents the output function of the PINN as parameterized by the network weights $\boldsymbol{w}$, and evaluated over a set of unlabeled collocation points sampled from the spatiotemporal domain. The weighting factors, $\lambda$, serve to balance the magnitude of different loss terms. The $N_{pde}$, $N_{ic}$, and $N_{bc}$ represent the number of training collocation points used for enforcing the PDE residual, initial conditions, and boundary conditions, respectively. The $\lVert \cdot \rVert _{L_2}$ represents the $L_2$ norm, which is used to quantify the magnitude of the loss terms.

\subsection{Integration of PINNs with numerical methods} \label{sec_pinn_nd}
While PINNs can exactly calculate the derivative terms in PDEs on a set of collocation points by AD, they fail to exploit potential properties that are both physically and numerically motivated \citep{karnakov2024solving}. These properties, common in well-established numerical methods, are advantageous for enhancing the overall model performance. In particular, the integral conservation property inherent in the FVM offers a way to reinforce the physical constraints within PINNs, maintaining its validity irrespective of grid resolution (\cite{patankar2018numerical}). Therefore, this study adopts FVM to exemplify its integration with PINNs, illustrating the potential benefits of such a hybridized approach.

\subsubsection{Hybrid framework combining PINN with FVM}

Taking the 2D steady lid-driven cavity flow problem presented in Section \ref{numerical experiments} as an example, the underlying physical laws are described by the 2D steady incompressible Navier–Stokes (N-S) equations:
\small
\begin{subequations} \label{eq:2dns}
    \begin{align}
        \frac{\partial u}{\partial x} + \frac{\partial v}{\partial y} &= 0 \label{eq:2dns-div} \\ 
         \frac{\partial \left(uu\right)}{\partial x} +  \frac{\partial \left(vu\right)}{\partial y} &= \frac{1}{Re}\left( \frac{\partial ^2u}{\partial x^2} + \frac{\partial ^2u}{\partial y^2} \right) - \frac{\partial p}{\partial x} \label{eq:2dns-m1} \\ 
       \frac{\partial \left( u v\right)}{\partial x} +  \frac{\partial \left(v v\right)}{\partial y} &= \frac{1}{Re}\left( \frac{\partial ^2v}{\partial x^2} + \frac{\partial ^2v}{\partial y^2} \right) - \frac{\partial p}{\partial y}  \label{eq:2dns-m2} 
    \end{align}
\end{subequations}
where $Re$ denotes the Reynolds number, a dimensionless quantity in fluid mechanics that can indicate the complexity of flow behavior. Let the dependent variable of interest within the momentum equations (Eqs. \ref{eq:2dns-m1} and \ref{eq:2dns-m2}) be designated as $\phi$. A general differential equation can then be formulated, encompassing the convection, diffusion, and source terms:
\small
\begin{equation} \label{gen_eq}
    \frac{\partial \left(u\phi\right)}{\partial x} + \frac{\partial \left( v\phi\right)}{\partial y} = \frac{1}{Re}\left( \frac{\partial ^2\phi}{\partial x^2} + \frac{\partial ^2\phi}{\partial y^2} \right) + S
\end{equation}
Here, the dependent variable $\phi$ represents the velocity components $u$ and $v$ in the $x$-direction  and $y$-direction, respectively. The source term $S$ corresponds to the pressure gradient, specifically  $- \frac{\partial p}{\partial x}$  for $x$-direction and $- \frac{\partial p}{\partial y}$ for $y$-direction, respectively.

\begin{figure}[ht]
    \centering
    \includegraphics[width=0.75\linewidth]{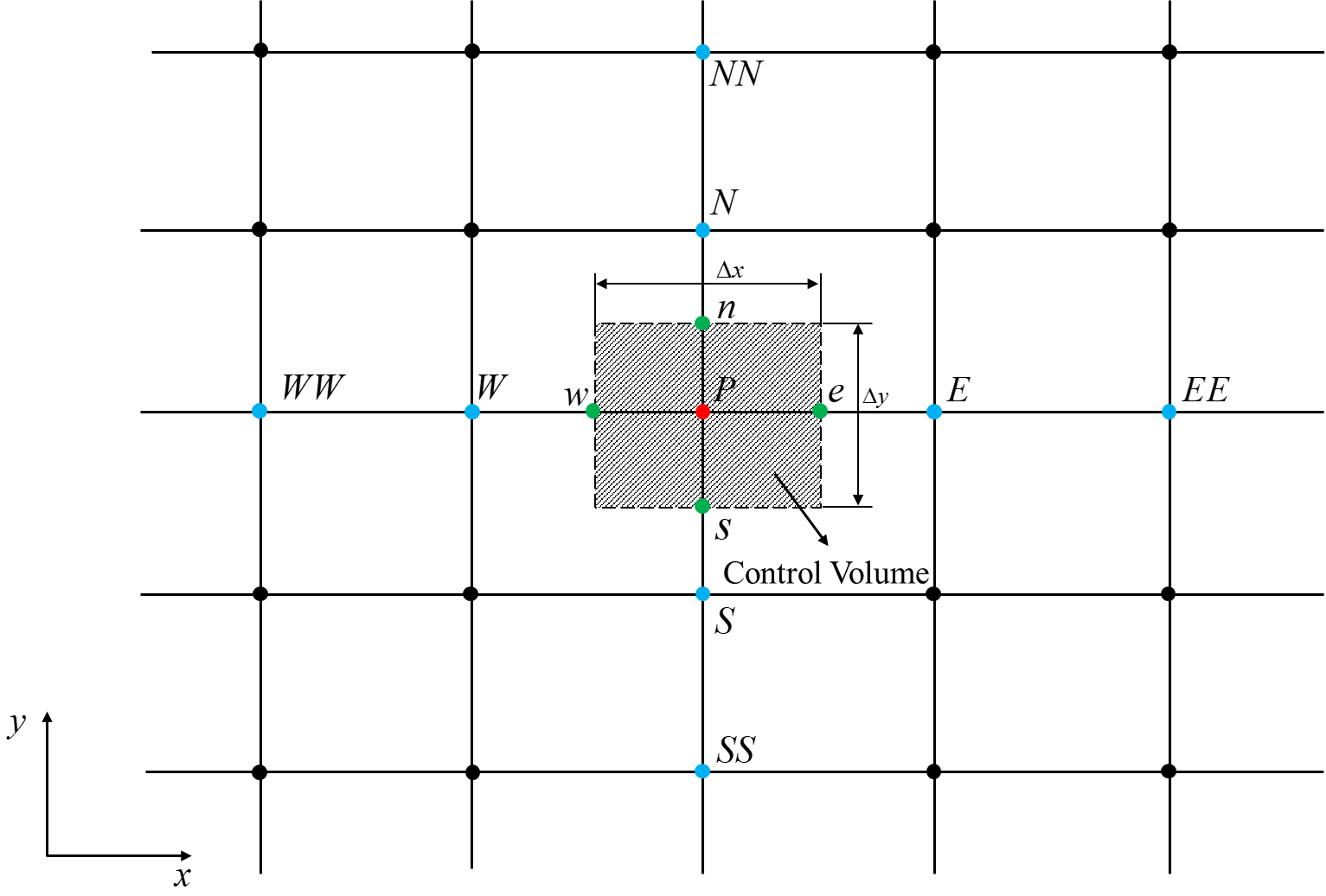}
    \caption{Control volume for the 2D domain. The capital letters (\textit{E}, \textit{W}, \textit{N}, \textit{S}, \textit{EE}, \textit{WW}, \textit{NN} and \textit{SS}) denote the surrounding nodes, while the lowercase letters (\textit{e}, \textit{w}, \textit{n} and $s$) represent the control surfaces. The capital letter \textit{P} refers to the center of the control volume. For simplicity of illustration, the spacing between adjacent grid points in both the \textit{x}-direction and the \textit{y}-direction is kept equal, with $\Delta x$ = $\Delta y$, but this is not a necessity.}
    \label{fig: cv}
\end{figure}

Performing a volume integral of Eq. \ref{gen_eq} over the control volume ($\Delta x$ = $\Delta y$) illustrated in Fig. \ref{fig: cv} yields the integral conservation form. This formulation embodies the fundamental principle of physical quantity conservation, applicable to quantities such as momentum or energy, within the defined control volume $V$:
\small
\begin{equation} \label{gen_eq_int}
\int_V \left (\frac{\partial \left(u\phi\right)}{\partial x} + \frac{\partial \left( v\phi\right)}{\partial y} \right )\, dV = \int_V\frac{1}{Re}\left( \frac{\partial ^2\phi}{\partial x^2} + \frac{\partial ^2\phi}{\partial y^2} \right) \, dV + \int_V S \, dV
\end{equation}

In the traditional FVM, the diffusion term is typically discretized using the central difference (CD) scheme, which offers a simple and effective method for approximating second-order spatial derivatives \citep{patankar2018numerical,malalasekera2007introduction,darwish2016finite}. The diffusion term in Eq. \ref{gen_eq_int} is discretized approximately as follows:
\small
\begin{equation} 
     \int_V \frac{1}{Re}\left( \frac{\partial ^2\phi}{\partial x^2} + \frac{\partial ^2\phi}{\partial y^2} \right)\, dV \approx    \frac{1}{Re}\left( \phi_E + \phi_W +\phi_N +\phi_S -4\phi_P \right)
\end{equation}
Subsequently, the source term in the \textit{x}-direction and \textit{y}-direction can be discretized as follows:
\small
\begin{subequations} 
\begin{align}
 &\int_V -\frac{\partial p}{\partial x} \, dV  \approx  \frac{-\left (p_E - p_W \right )\Delta y}{2} \\
 &\int_V -\frac{\partial p}{\partial y} \, dV  \approx  \frac{-\left (p_N - p_S \right )\Delta x}{2} 
 \end{align}
\end{subequations}
Furthermore, volume integration of the convection term is carried out:
\small
\begin{equation} \label{con_int}
   \int_V \frac{\partial \left(u\phi\right)}{\partial x} \, dV = \int_s^n \int_w^e \frac{\partial \left(u\phi\right)}{\partial x} dx \, dy  \approx \left [ (u\phi)_e - (u\phi)_w \right ]\Delta y
\end{equation}
We introduce a new symbol F, defined as  \( F = u \Delta y \), which represents the strength of the convection. Hence,
Eq. \ref{con_int} can  be expressed as:
\small
\begin{equation} \label{con_int_2}
    \left [ (u\phi)_e - (u\phi)_w \right ]\Delta y =  (F\phi)_e - (F\phi)_w 
\end{equation}

A variety of discretization schemes are available for approximating the convection term.
Taking $\phi_e$ as example,
the following discretization can be derived:
\begin{itemize}
    \item \textit{\textbf{Central difference (CD) scheme}}
\end{itemize}
\begin{align}
\phi_e = \frac{\phi_E + \phi_W}{2}
\end{align}
\begin{itemize}
    \item \textit{\textbf{First order upwind (FOU) scheme}}
\end{itemize}
\begin{align}
\phi_e = 
\begin{cases}
 \phi_P, & F_e \ge 0\\
 \phi_E, & \text{otherwise}
\end{cases}
\end{align}
\begin{itemize}
    \item \textit{\textbf{Second order upwind (SOU) scheme}}
\end{itemize}
\begin{align}
\phi_e = 
\begin{cases}
 \frac{3}{2}\phi_P - \frac{1}{2}\phi_W, & F_e \ge 0\\
 \frac{3}{2}\phi_E - \frac{1}{2}\phi_{EE}, & \text{otherwise}
\end{cases}
\end{align}

By substituting the aforementioned discretized forms of the convection term, diffusion term, and source term into Eq. \ref{gen_eq_int}, we derive the following general discretized equation:
\begin{equation} \label{disc_eq}
{\textstyle a_{P}\phi_{P} + \sum a_{NB}\phi_{NB}+ b} = 0
\end{equation}
where  $a_P$  is the coefficient associated with the central node  $P$ , $ a_{NB}$  represents the coefficients of the neighboring nodes  $NB$ (comprising $E$, $W$, $N$, $S$, $EE$, $WW$, $NN$, and $SS$), and $b$ is the source term.  % The detailed expressions for the coefficients and the term $b$ are provided in  \ref{sec: appendix coefficient fvm}. 
Within the framework that integrates PINNs with the FVM, the PDE-based physical loss ($L_{pde}$ ) is transformed into a physical loss based on the discretized equation ($L_{de}$). Specifically, the physical loss term imposed by the momentum equation is expressed as follows:
\small
\begin{equation} \label{eq:physics_loss_sou_de}
    \mathcal{L}_{de} = \frac{1}{N_{de}}\lVert  a_{P}\phi_{P} + \sum a_{NB}\phi_{NB}+ b \rVert _{L_2(\Omega \times (0,T])}^2
\end{equation} 

Comparing Eq. \ref{eq:physics_loss_sou_de} with Eq. \ref{PDE_loss}, it can be observed that the loss term based on the discretized equation establishes an algebraic relationship between the collocation point and its surrounding nodes, enabling a more effective imposition of physical constraints.
\subsection{The FFV-PINN framework}
\subsubsection{The simplified FVM}
A variety of discretization schemes exist for the convection term, including FOU, CD, and SOU, among others. The appropriate choice of discretization scheme typically depends on the direction of the convection or the relative magnitudes of convection and diffusion strength.
% , quantified by the Peclet number. 
In convection-dominated flows, upwind schemes are frequently favored to ensure numerical stability and capture the directional nature of convective transport. Conversely, in diffusion-dominated flows, 
the CD scheme is generally considered more suitable. It is therefore apparent that the selection of a discretization scheme for the convection term is a non-trivial task that requires a thorough understanding of the underlying physical phenomena and the strengths and limitations of different numerical methods. 

Given the complexities associated with this decision-making process, a more robust discretization scheme for the convection term is needed in the context of PINNs. To this end, we introduce a simplified FVM, which leverages the neural network's ability to produce outputs at arbitrary positions. The discretized form of the convection term denoted as Eq. \ref{con_int}, can then be implemented directly without requiring further discretization for cell-face velocity. As a result, the integral form of convective term $\frac{\partial \left(v\phi\right)}{\partial y}$, when discretized in the control volume depicted in Fig. \ref{fig: cv}, results in the following expression:
\small
\begin{equation} \label{convec_v}
   \int_V \frac{\partial \left(v\phi\right)}{\partial y}\, dV   \approx \left [ (v\phi)_n - (v\phi)_s \right ]\Delta x
\end{equation}

The simplified FVM enhances the flexibility and efficiency of the discretization process, mitigating the inherent complexities typically associated with traditional FVM. Moreover, the proposed scheme exhibits superior dispersion and dissipation properties based on theoretical analysis detailed in Section \ref{02_04}.

By systematically organizing the discretized forms of the convection term, diffusion term, and source term, we arrive at the following expression:
\small
\begin{equation}\label{eq_dis}
 a_{P}\phi_{P} + \sum a_{NB}\phi_{NB} +\sum a_{nb}\phi_{nb} + b = 0
\end{equation}
where the $a_P$ represents the coefficient associated with the central node $P$, $a_{NB}$ denotes the coefficients corresponding to the neighboring nodes $NB$ (each of $E$, $W$, $N$, $S$) surrounding the given control volume, and  $a_{nb}$ refers to the coefficients associated with the control faces $nb$ ($e$, $w$, $n$, $s$) of a given control volume. The detailed coefficients can be found in \ref{sec: appendix coefficient fvm}. Therefore, the physical constraints derived from the momentum equation can be re-formulated as follows:
\small
\begin{equation} \label{eq:physics_loss_face_de}
    \mathcal{L}_{de} = \frac{1}{N_{de}}\lVert a_{P}\phi_{P} + \sum a_{NB}\phi_{NB}+\sum a_{nb}\phi_{nb} + b \rVert _{L_2(\Omega \times (0,T])}^2
\end{equation} 

It is worth noting that, through discretization, conservation laws are naturally incorporated into the loss function in the form of physics-informed constraints. Theoretically, if the training points, which are uniformly distributed, fully cover the computational domain and the loss function converges to zero, FFV-PINN can exactly satisfy global conservation laws. In practice, however, mini-batch training is adopted due to computational limitations, leading to the local enforcement of conservation. Although conservation is enforced only at the local level, these constraints cumulatively lead to an accurate approximation of global conservation as training progresses and the loss decreases. The subsequent numerical experiments in Section \ref{numerical experiments} further verify that FFV-PINN outperforms traditional AD-based PINNs in terms of both accuracy and computational efficiency, thereby demonstrating its effective compatibility with conservation laws.

\subsubsection{Residual correction loss}

Having established the physical constraints based on FVM, it is now pertinent to highlight the fundamental differences in how PINNs and FVM obtain the solution of PDEs. Whereas FVM relies on constructing a linear system to solve PDEs directly or iteratively, PINNs instead provide a markedly distinct paradigm by recasting a numerical problem as an optimization problem. Specifically, the neural network adjusts its parameters by minimizing a defined loss function, with the objective of reducing each loss term as close to zero as possible. This optimization constitutes the core of the neural network’s training process. Consequently, to enforce compliance with the governing equations, PINNs are designed to minimize the physics-based loss term, defined in Eq. \ref{eq:physics_loss_face_de}, with the ideal target of achieving zero.  However, it is critical to acknowledge the distinction between theoretical ideal and practical reality. While perfect convergence, in theory, would indeed reduce this term to zero, practical implementations invariably yield finite residuals, denoted as $r^{n}$ at the training epoch $n$. The residuals quantitatively represent the extent of the current solution's deviation from the governing equations, and this is formalized as follows:
\small
\begin{equation} \label{eq:residual_de_n}
a_{P}\phi _{P}^{n} +\sum{a_{NB}}\phi _{NB}^{n} + \sum{a_{nb}^{n}}\phi _{nb}^{n}+{{b}^{n}}=r^{n}
\end{equation} 

The coefficients \(a_P\) and \(a_{NB}\) are predetermined based on the discretization scheme and the physical properties of the system as detailed in Eq.\ref{eq_coff}. Critically, these coefficients remain constant and unaffected by neural network parameter updates. Let us now assume the physical constraints are met in the subsequent update step—the $(n+1)$-th epoch. Under this assumption, the physics loss is expressed as follows:
\small
\begin{equation} \label{eq:residual_de_n+1}
a_{P}\phi _{P}^{n+1}+\sum{a_{NB}}\phi _{NB}^{n+1}+ \sum{a_{nb}^{n+1}}\phi _{nb}^{n+1}+{{b}^{n+1}}=0
\end{equation} 
By subtracting Eq. \ref{eq:residual_de_n} from Eq. \ref{eq:residual_de_n+1}, the subsequent expression is derived:
\small
\begin{multline}  
\label{eq:residual_minus}
\sum \left(a_{nb}^{n+1}\phi_{nb}^{n+1} - a_{nb}^{n}\phi_{nb}^{n} \right)  
+ \sum a_{NB} \left(\phi_{NB}^{n+1} - \phi_{NB}^{n} \right) \\  
+ a_{P} \left(\phi_{P}^{n+1} - \phi_{P}^{n}\right)  
+ \left(b^{n+1} - b^{n}\right) = -r^{n}
\end{multline}
By rearranging Eq. \ref{eq:residual_minus}, we can obtain:
\small
\begin{equation}  
\label{eq:residual_frac}
\textstyle
 \phi _{P}^{n+1}- \phi _{P}^{n} = \frac{-r^{n} - \sum \left(a_{nb}^{n+1}\phi_{nb}^{n+1} - a_{nb}^{n}\phi_{nb}^{n} \right) -\sum{a_{NB}}\left(\phi _{NB}^{n+1} -\phi _{NB}^{n} \right) - \left( {{b}^{n+1}}-{{b}^{n}} \right)}{a_{P}} 
\end{equation}

When the update of $\phi_{P}^{n}$ is equal to Eq. \ref{eq:residual_frac}, this implies the physical constraints within the neural network are satisfied. The adjustments applied to the predictions are not arbitrary but systematically derived from discrepancies identified in the current prediction. This relationship is essential for ensuring that the model learns effectively and converges to a solution that adheres to the physical laws governing the system. 

Inspired by the SIMPLE algorithm~\citep{patankar2018numerical} in the FVM, we propose a modification to Eq.~(\ref{eq:residual_frac}) by neglecting the term $\sum \left(a_{nb}^{n+1}\phi_{nb}^{n+1} - a_{nb}^{n}\phi_{nb}^{n} \right)$. This simplification facilitates implementation and improves model efficiency. As a result, we obtain the following approximation:

\small
\begin{equation} \label{eq:residual_}
\phi _{P}^{n+1}- \phi _{P}^{n} \approx R = \frac{-r^{n}-\sum{a_{NB}}\left(\phi _{NB}^{n+1} -\phi _{NB}^{n} \right)-\left(b^{n+1}-b^{n}\right)}{a_{P}}
\end{equation} 
where $R$ is defined as the residual correction term. As discussed in the 
SIMPLE algorithm \citep{patankar2018numerical}, neglecting the contributions from neighboring nodes may lead to overestimating the correction terms. Therefore, to mitigate this issue and enhance the stability of the training process, we introduce $\alpha$, a relaxation factor for the residual correction to modulate the influence of $R$, leading to the following expression:
\small
\begin{equation}\label{cor_phi}
\phi_{P}^{n+1}  = \phi_{P}^{n} + \alpha R
\end{equation}

To correct the update behavior of the neural network, we define the residual correction loss term $L_{rc}$ using the $L_1$ norm as:

\begin{equation} \label{cor_phi_loss} L_{rc}= \frac{1}{N_{rc}} \left| \phi^{n+1}_{P} - \phi^{n}_{P} - \alpha R \right|_{L_1(\Omega \times (0,T])} \end{equation}

Since the value at the $(n+1)$-th epoch is not directly available during training, we estimate it using a second-order extrapolation scheme, given by:
\small
\begin{subequations} \label{assume}
\begin{align}
 & \phi^{n+1}=2\phi ^{n}-\phi ^{n-1} \label{assume_phi} \\
 & b ^{n+1}=2b^{n}-b ^{n-1} \label{assume_b}
 \end{align}
\end{subequations}
Substituting Eq.~\eqref{assume} into Eq.~\eqref{cor_phi_loss}, we simplify the loss term as:
\begin{equation} \label{cor_phi_loss_simplified} 
L_{rc} = \frac{1}{N_{rc}} \left| \phi^{n}_{P} - \phi^{n-1}_{P} - \alpha R \right|_{L_1(\Omega \times (0,T])}
\end{equation}
where $R$ is now expressed as:
\begin{equation} \label{eq:residual_linear}
R = \frac{-r^{n}-\sum{a_{NB}}\left(\phi _{NB}^{n} -\phi _{NB}^{n-1} \right)-\left(b^{n}-b^{n-1}\right)}{a_{P}}
\end{equation}

% Through algebraic simplification, we obtain:
% \begin{equation} \label{cor_phi_loss_final} L_{rc} = \frac{1}{N_{rc}} \left| \phi^{n}_{P} - \phi^{n-1}_{P} - \alpha R \right|_{L_1(\Omega \times (0,T])} \end{equation}

% Upon substitution of Eq. \ref{assume} into Eq. \ref{eq:residual_}, the subsequent expression is obtained:

% Based on Eq.\ref{assume_phi} and Eq.\ref{cor_phi}, a residual correction loss term ($L_{rc}$) is defined using $L_1$ norm, as follows:
% \small
% \begin{equation}\label{cor_phi_loss}
% L_{rc}= \frac{1}{N_{rc}}\left| \phi^{n+1}_{P} -\phi^{n}_{P}-\alpha R  \right|_{L_1(\Omega \times (0,T])}=\frac{1}{N_{rc}}\left| 2 \phi^{n}_{P} -\phi^{n-1}_{P} -\phi^{n}_{P}-\alpha R  \right|_{L_1(\Omega \times (0,T])}=\frac{1}{N_{rc}}\left| \phi^{n}_{P} -\phi^{n-1}_{P}-\alpha R  \right|_{L_1(\Omega \times (0,T])}
% \end{equation}

% Given that the variable at the future time step $n+1$ is unknown, we employ a well-established second-order extrapolation method to estimate it. T
The optimal choice of $\alpha$ is inherently problem-dependent and lacks a universal selection criterion~\citep{OpenFOAMFVM}. It is typically determined through empirical tuning and is influenced by factors such as the problem type, neural network architecture, and the number of collocation points. Although $\alpha$ can be individually tailored for different variables, for simplicity and to minimize parametric complexity in the FFV-PINN framework, the same relaxation factor is applied to all variables. 

It is noted that the newly proposed residual correction loss term plays a crucial role in directing the training trajectory of the neural network.  It theoretically guarantees that updates are executed in a manner concordant with the fulfillment of physical constraints. The underlying rationale for establishing the residual correction is that $R$ quantifies the difference between predicted values and perfect converged, physics-consistent outcomes, as estimated by the last two epochs.
% By setting the update step size of $\phi_{P}$ to $R$, the training process directly corrects the residual determined by the model.

In summary, the proposed FFV-PINN framework, as illustrated in Fig. \ref{fig:nn}, incorporates two essential losses in addition to the BC loss: the physical constraint loss (Eq. \ref{eq:physics_loss_face_de}), which ensures adherence to physical laws, and the residual correction loss (Eq. \ref{cor_phi_loss}), which minimizes model residuals to enhance convergence and stability.  
% In addition to BC loss, it comprises two essential components, each serving a distinct purpose: physical constraint loss (Eq. \ref{eq:physics_loss_face_de}) to ensure adherence to physical laws, and residual correction loss (Eq. \ref{cor_phi_loss}) to minimize model residual for converge and stability.
These loss terms are carefully designed to enhance model performance, ensuring predictive accuracy via adherence to the underlying physical principles governing the system. 

\subsection{Fundamental analysis}\label{02_04}
We further analyze the dispersion and dissipation properties of the proposed discretization scheme for the convection term. To this end, Fourier analysis is employed. Assuming the field variable is denoted as $\phi$, its Fourier transform and inverse transform can be expressed as follows.
\begin{itemize}
    \item \textbf{\textit{Fourier Transform:}}
\end{itemize}
\small
\begin{equation} \label{fourier}
\hat{\phi}(k) = \frac{1}{2\pi}\int_{-\infty}^{\infty} \phi(x) e^{-ikx} \, dx  
\end{equation}

\begin{itemize}
    \item \textbf{\textit{Inverse Fourier Transform:}}
\end{itemize}
\small
\begin{equation} \label{inverse_fourier}
\phi(x) = \frac{1}{2\pi} \int_{-\infty}^{\infty} \hat{\phi}(k) e^{ikx} \, dk 
\end{equation} 
where $k$ represents the wavenumber, and $x$ is the spatial coordinate, $i$ denotes the imaginary unit. For the simplified FVM, the spatial derivative $\frac{\partial \phi}{\partial x} $ can be mathematically represented as follows:
\small
\begin{equation}\label{partial_spatial}
    \frac{\partial \phi}{\partial x} = \frac{\phi_e - \phi_w}{h} = \frac{\phi (x + \frac{h}{2}) - \phi (x - \frac{h}{2})}{h}
\end{equation}
where $h =\Delta x$, and $\phi_e$ and $\phi_w$ are obtained directly by the neural network.

Subsequently, the Fourier transform is applied to Eq. \ref{partial_spatial}.
\small
\begin{equation}
    ikh\hat{\phi}(k) =  e^{i\frac{kh}{2}} \hat{\phi}(k) - e^{-i\frac{kh}{2}} \hat{\phi}(k) 
\end{equation}
Through manipulation of the aforementioned equation, the following is derived:
\small
\begin{equation}
    ikh = e^{i\frac{kh}{2}}  - e^{-i\frac{kh}{2}} 
\end{equation}
Under the approximation that the effective wavenumber $k^{'}  \cong  k$, and multiplying both sides of the above equation by $-i$, we can deduce: 
\small
\begin{equation}\label{kh}
    k^{'}h = -i\left( e^{i\frac{kh}{2}}  - e^{-i\frac{kh}{2}} \right)
\end{equation}
The real and imaginary parts of Eq. \ref{kh} can be further simplified as:
\small
\begin{subequations}
\begin{align}
& \Re \left( k^{'}h \right) = \Re \left[ - i \left( e^{i\frac{kh}{2}} - e^{-i\frac{kh}{2}} \right) \right] = 2 \sin\left(\frac{kh}{2}\right) \\
& \Im \left( k^{'}h \right) =\Im \left[ - i \left( e^{i\frac{kh}{2}} - e^{-i\frac{kh}{2}} \right) \right] = 0
\end{align}
\end{subequations}
The dispersion ($\kappa_i$) and dissipation ($\kappa_r$) can be defined as:
\small
\begin{subequations}
\begin{align}
& \kappa_i = \Re \left( k^{'}h \right)  \\
& \kappa_r = \Im \left( k^{'}h \right)
\end{align}
\end{subequations}

\begin{figure}
    \centering
    \includegraphics[width=1\linewidth]{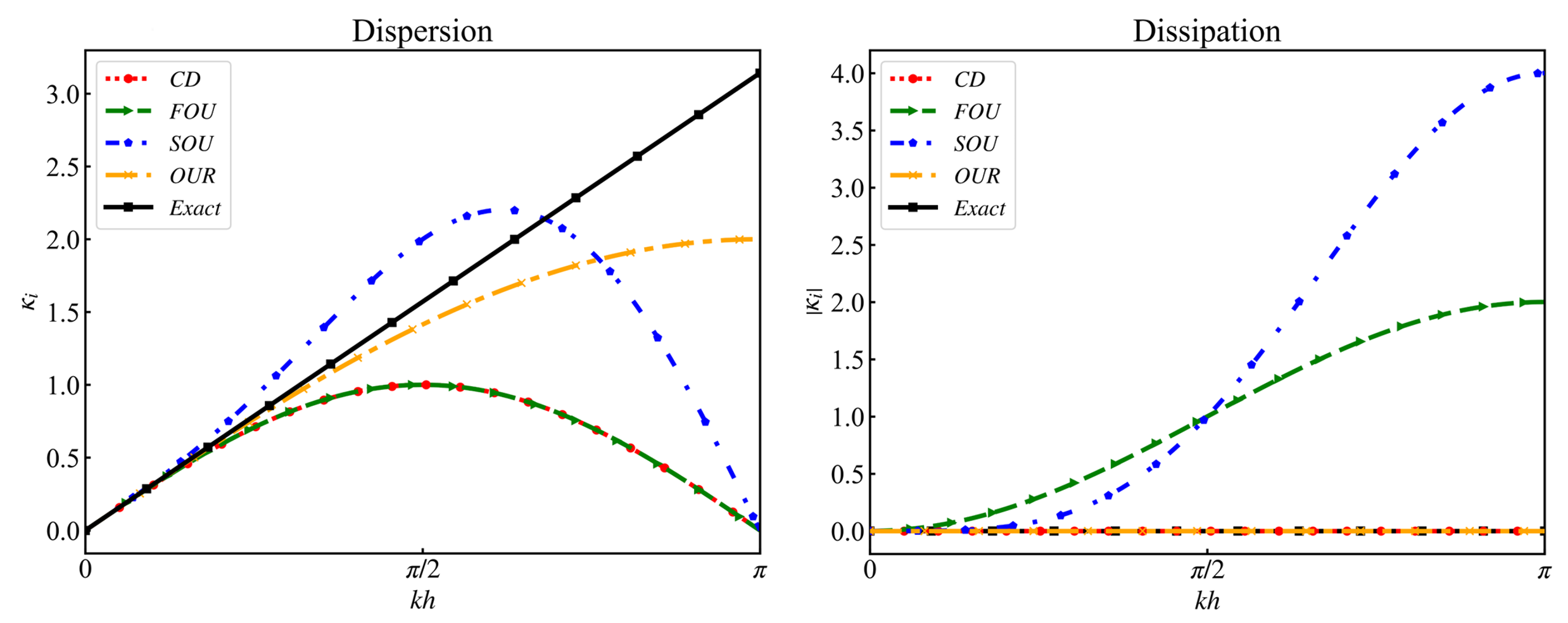}
    \caption{Comparison of the dispersion and dissipation characteristics of each scheme.}
    \label{fig: fundamental analysis}
\end{figure}

Fig. \ref{fig: fundamental analysis} compares the proposed scheme with established schemes (CD, FOU, and SOU) by analyzing the relationship between $kh$, $\kappa_i$ and $ \left |\kappa_r \right |$. This comparison also yields insight into the fundamental dispersion and dissipation characteristics of each scheme. A detailed fundamental analysis of the established schemes is provided in \ref{sec: appendix analysis}.
As demonstrated in Fig. \ref{fig: fundamental analysis}, the proposed scheme closely matches the exact solution in terms of dispersion. It outperforms other schemes like FOU and SOU, which exhibit significant deviations from the exact solution, especially in the high-frequency range. Another key advantage of the proposed scheme is its zero-dissipation property. This non-dissipative characteristic ensures that higher frequency components of the solution can be better captured. Consequently, the fundamental analysis suggests that the developed scheme exhibits superior numerical properties compared to existing methods, even without considering its simpler implementation. 

\section{Numerical experiments} \label{numerical experiments}

In this section, we conduct a series of numerical experiments, including 2D steady and unsteady lid-driven cavity flows, 3D steady lid-driven cavity flow, backward-facing step flow, and natural convection in a unit square. The selected benchmark problems are well-known in the fluid dynamics community for their complex nonlinear dynamics and challenging characteristics. By conducting these experiments, we illustrate the advantages of FFV-PINN in terms of stability, convergence, and accuracy relative to other PINNs.

To enhance the network's capability in capturing high-frequency information, a frequency mapping layer is introduced, where the input coordinates $(x,y)$ are transformed into a high-dimensional space, as illustrated in Fig. \ref{fig:nn}. For unsteady problems, the temporal coordinate $t$ is included, and the mapping is performed on the input coordinates $(t, x, y)$. Similarly, for 3D problems, the spatial coordinate $z$ is considered, and the mapping is applied to the coordinates $(x, y, z)$. Furthermore, to augment the network's expressive capacity, the architecture employs both shared and variable-specific hidden layers.  Specifically, two shared hidden layers are utilized, followed by three variable-specific hidden layers dedicated to the velocity components $u$, $v$, and pressure $p$, respectively. Each hidden layer consists of 32 neurons and employs the SiLU (Sigmoid Linear Unit) function as the activation function. In 3D problems and natural convection problems, additional output branches are incorporated to represent the velocity component $w$ and the temperature field $T$, respectively. The neural network is trained using the Adam optimizer, with the learning rate following a warmup cosine decay schedule. All implementations are developed using the JAX library and executed on a single NVIDIA RTX 3090 GPU card. 

Table \ref{tab:fvv-pinn_training}  presents a list of additional parameter configurations used in the numerical experiments, including details on batch size, maximum iterations, initial learning rates, weights for each loss term, relaxation factors, and warmup steps. It should be emphasized that all parameters listed in the table were determined empirically or through a process of trial and error, without the application of rigorous hyperparameter optimization techniques. The primary objective of this study is to demonstrate the validity and advantages of the proposed method itself, as opposed to the exhaustive optimization of hyperparameters to achieve optimal performance.

In the remaining numerical experiments presented in this study, $\alpha$ was varied between 0.7 and 0.98. For real-world systems where the true solution remains unknown, we recommend determining $\alpha$  by carefully monitoring the loss function during the training process. Specifically, an appropriate value of $\alpha$ typically manifests as a stable and consistent decrease in the loss function. Conversely, an unsuitable $\alpha$  may lead to oscillations, slow convergence, or even divergence of the loss. Consequently, observing the trajectory of the loss function provides a crucial empirical indicator for assessing the suitability of a chosen $\alpha$ for a given problem. Furthermore, the emergence of meta-learning techniques offers a promising avenue for automating this search \citep{zoph2016neural,finn2017model}, a direction we intend to explore in our future research.

\begin{table}[!ht]
\centering
\caption{
Training configurations used in the numerical experiments for five benchmark problems: 2D steady lid-driven cavity (LDC) flow ($Re = 5000$), 2D unsteady LDC flow ($Re = 400$), 3D steady LDC flow ($Re = 1000$), backward-facing step (BFS) flow ($Re = 600$), and natural convection (NC) problem ($Ra = 10^8$). The batch size is calculated as $k_{de} \times N_{de}$, $k_{bc} \times N_{bc}$, and $k_{ic} \times N_{ic}$ (for unsteady cases), where $N_{de}$, $N_{bc}$, and $N_{ic}$ represent the number of training points used for physical constraints, boundary conditions, and initial conditions, respectively. The coefficients $k_{de}$, $k_{bc}$, and $k_{ic}$ denote the corresponding sampling ratios. Other key parameters include the maximum iterations (Max iter.), initial learning rate (Init. LR), and relaxation factor (Relax. factor).
}

\label{tab:fvv-pinn_training}
\vspace{0.5em} % Small vertical space below caption
\renewcommand{\arraystretch}{1.2} % Good balance for row spacing
\small % A good size for readability without being overwhelming
\resizebox{\textwidth}{!}{
\begin{tabular}{ccccccc}
\toprule
\multirow{3}{*}{Parameter} & \multicolumn{3}{c}{LDC} & BFS & NC\\
 \cmidrule(lr){2-4}
& 2D steady & 2D unsteady & 3D steady &\multirow{2}{*}{$Re = 600$} &\multirow{2}{*}{$Ra = 10^8$}\\
 & $Re = 5000$ & $Re = 400$ & $Re = 1000$ & & \\
\midrule
% \multirow{3}{*}{Mesh size} &  $\Delta x =\frac{L_x}{N_x}$ & $\Delta x =\frac{L_x}{N_x}$ & $\Delta x =\frac{L_x}{N_x}$ & $\Delta x = \frac{L_x}{N_x}$ & $\Delta x = \frac{L_x}{N_x}$  \\
%  &  $\Delta y =\frac{L_y}{N_y}$ & $\Delta y =\frac{L_y}{N_y}$ & $\Delta y =\frac{L_y}{N_y}$ & $\Delta y =\frac{L_y}{N_y}$ & $\Delta y =\frac{L_y}{N_y}$  \\
% &  -  &  - & $\Delta z =\frac{L_z}{N_z}$ & -  &  - \\
\addlinespace % Adds a little more vertical space between groups of rows

\multirow{3}{*}{Batch size} & $0.1\times N_{de}$ & $0.01\times N_{de}$ & $0.001\times N_{de}$ & $0.1\times N_{de}$ & $0.1\times N_{de}$  \\
  &$ 0.1\times N_{bc}$ & $0.05\times N_{bc}$ & $0.001\times N_{bc}$ & $0.1\times N_{bc} $ & $0.1\times N_{bc}$ \\
  & - & $0.01\times N_{ic}$ & -  &  - & - \\
\addlinespace
% \cmidrule(lr){2-6}

% \addlinespace
\cmidrule(lr){2-6}
\multirow{4}{*}{Weights ($\lambda_{i}$)} & $\lambda_{de} = 1 $ &$\lambda_{de} = 1 $ & $\lambda_{de} = 1 $ &$\lambda_{de} = 8 $ & $\lambda_{de} = 1 $  \\
 & $ \lambda_{bc} = 1$ &$ \lambda_{bc} = 1 $ & $ \lambda_{bc} = 1 $ &$ \lambda_{bc} =  1 $ & $ \lambda_{bc} = 1 $  \\
  & - &$ \lambda_{ic} = 1 $ &- & - &  -   \\
 & $\lambda_{rc} = 5 $ &$\lambda_{rc} = 10 $ & $\lambda_{rc} = 1 $ &$\lambda_{rc} = 1 $ & $\lambda_{rc}  = 1 $  \\
\cmidrule(lr){2-6}

 Max iter. & 50000 & 50000 & 100000 & 50000 & 50000  \\
\addlinespace
Init. LR & 1e-3 & 1e-3 & 1e-3 & 5e-3 & 5e-3  \\
Relax. factor & 0.79 & 0.94 & 0.8 & 0.97 & 0.84  \\
\addlinespace
Warmup steps & 0 & 0 & 0 & 7500 & 15000  \\
% \addlinespace
\bottomrule
\end{tabular}
}
\end{table}

To quantitatively evaluate the accuracy of the FFV-PINN, this study utilizes two primary metrics: the mean squared error (MSE) and the relative $L_2$ error. These metrics are essential in quantifying the discrepancies between the model's predicted solutions and the corresponding numerical (or benchmark) solutions. The MSE is defined as the average of the square of differences between the predicted and actual values. Mathematically, it is expressed as:
\begin{equation} \label{eq:mse}
\text{MSE} = \frac{1}{N} \sum_{i=1}^{N} (y_i - \hat{y}_i)^2
\end{equation} 
where $y_i$ represents the actual values from the numerical solution, $\hat{y}_i$ denotes the predicted values from the FFV-PINN model, and $N$ is the total number of data points.  

The relative $L_2$ error provides a normalized measure of the discrepancy between the predicted and actual values, facilitating an assessment of the model's performance relative to the magnitude of the true solution. Mathematically, it is defined as:
\begin{equation} \label{eq:l2error}
\text{Relative $L_2$ error} = \frac{\| y - \hat{y} \|_{L_2}}{\| y \|_{L_2}}
\end{equation} 
where $\| y - \hat{y} \|_2$ represents the $L_2$ norm of the error vector, and $\| y \|_2$ is the $L_2$ norm of the actual values.
\subsection{Lid-driven cavity flow}
\begin{figure}[!ht]
    \centering
    \includegraphics[width=0.5\linewidth]{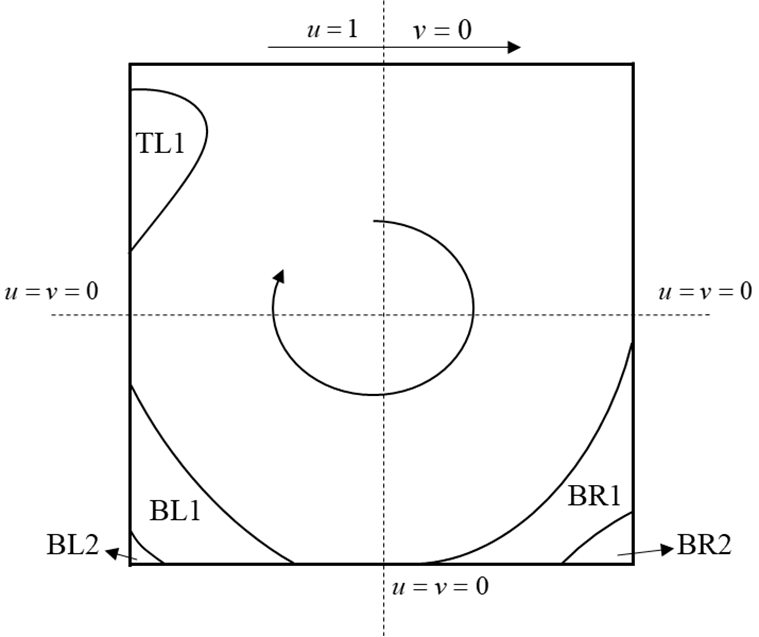}
    \caption{Schematic of the lid-driven cavity problem with key locations marked (TL1, BL1, BL2, BR1, BR2). The top lid moves at a constant velocity ($u=1,v=0$), while the other walls are stationary ($u=v=0$). The central region indicates the expected formation of a recirculating vortex.}
    \label{fig:ldc}
\end{figure}

The lid-driven cavity flow problem has become a benchmark for the validation of emerging numerical methods in computational fluid dynamics, particularly following the seminal work by \citep{ghia1982high}. The problem is defined within a geometrically simple yet physically rich domain: a 2D $1 \times 1$ unit square cavity. The fluid is driven by the tangential motion of the top wall, which moves at a constant horizontal velocity of $u_{\text{lid}} = 1$ while maintaining a zero vertical velocity component ($v_{\text{lid}} = 0$), as schematically illustrated in Fig. \ref{fig:ldc}. In addition, a no-slip boundary condition is imposed on the remaining three walls of the cavity. The linear local structure approximation method \citep{wong2023lsa} is employed for boundary conditions in this study, except where otherwise noted. The physics laws governing the system are described by the incompressible N-S equations, as expressed in Eq. \ref{eq:2dns}.

The simulation of lid-driven cavity flow at high $Re$ (e.g., $Re$ = 3200, 5000) with PINN presents significant challenges. As highlighted in recent studies \citep{wang2023expert,wang2023solution,wang2024piratenets,cao2023tsonn}, PINNs often struggle with training instability and accuracy issues when applied to such problems. Despite these difficulties, our proposed FFV-PINN demonstrates a remarkable ability to maintain both fast convergence and accuracy, as shown in Table \ref{ldc_table}. Table \ref{ldc_table} also summarizes the performance of various PINNs on this benchmark ($Re$ = 3200, 5000). Notably, our proposed method outperforms existing approaches, achieving the lowest relative $L_2$ error. More critically, in comparison to the state-of-the-art SOAP method \citep{2025arXivSOAP}, our approach attains a remarkable reduction in computational cost: while employing a less powerful NVIDIA RTX 3090 GPU (versus the A6000 utilized by SOAP), FFV-PINN achieves a computational speed that is $125 \times$ faster than SOAP. These results highlight the FFV-PINN framework’s dual strengths of computational efficiency and accuracy in simulating complex fluid dynamics, even under challenging high $Re$ regimes.
\begin{table}[ht]
\centering
\caption{Comparison of relative $L_2$ errors (computed for the velocity components $u$, $v$, and the velocity magnitude $V$) and training time (in \textit{hour}) obtained by different PINNs variants. The computation times of SOAP are provided in the respective paper, while the training time for PirateNet was obtained by running the authors' provided code on an NVIDIA RTX 3090 GPU card. Additionally, to ensure consistency, we calculated the relative $L_2$ errors using the benchmark data provided by PirateNet.}
\label{ldc_table}
\resizebox{\textwidth}{!}{

\begin{tabular}{ccccccc}
\toprule
References & $Re$ &  $u$ & $v$ & $V$ & Time & Hardware\\
\midrule
JAXPI \citep{wang2023expert} &3200 & - & - & 1.58e-01 & -& -\\
PirateNet \citep{wang2024piratenets}& 3200 & -  & - & 4.21e-02 & 11.83h
&RTX 3090\\
ev-NSFnet \citep{wang2023solution}& 5000 & 5.40e-02 & 5.40e-02 & - & -&- \\
TSONN \citep{cao2023tsonn}& 5000 & - & - & 1.00e-01 & -&-\\
SOAP \citep{2025arXivSOAP}& 5000 & - & - & 3.99e-02 &  8.25h& RTX A6000\\
Ours & 3200 & 3.78e-02 & 3.72e-02 & 3.69e-02 & 0.0636h& RTX 3090\\
Ours & 5000 & 4.05e-02 & 4.05e-02 & 3.90e-02 & 0.0658h& RTX 3090\\
\bottomrule
\end{tabular}
}
\end{table}

\begin{figure}[!ht]
    \centering
    \includegraphics[width=1.0\linewidth]{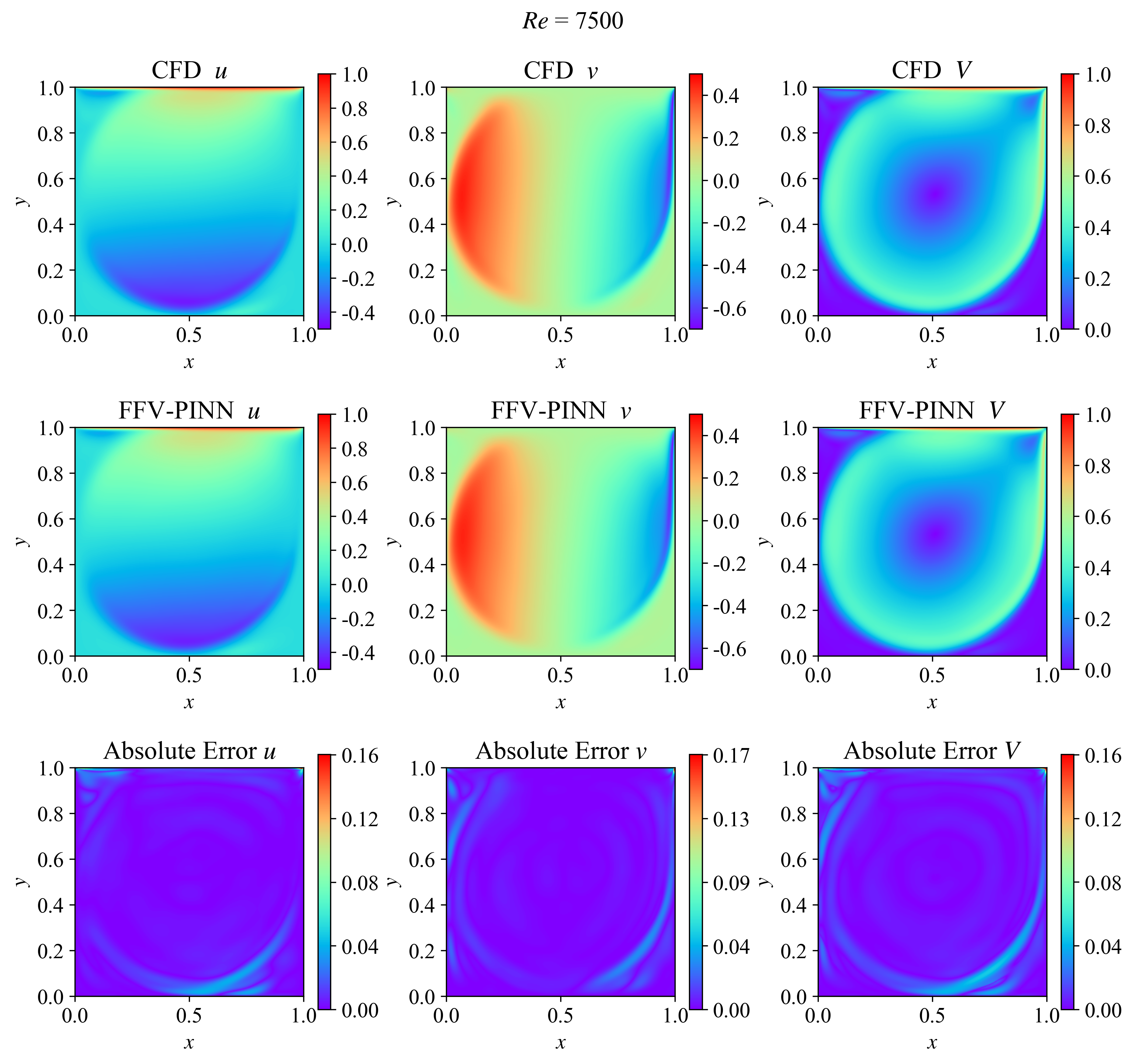}
    \caption{Lid-driven cavity flow at $Re$ = 7500. The left column represents the velocity component $u$, the middle column corresponds to the velocity component $v$, and the right column depicts the velocity magnitude $V$. Each row corresponds to, respectively, the solutions obtained via CFD, the proposed FFV-PINN method, and the absolute error between the two.}
    \label{fig:ldc_7500_contour}
\end{figure}

\begin{figure}[!ht]
    \centering
    \includegraphics[width=1.0\linewidth]{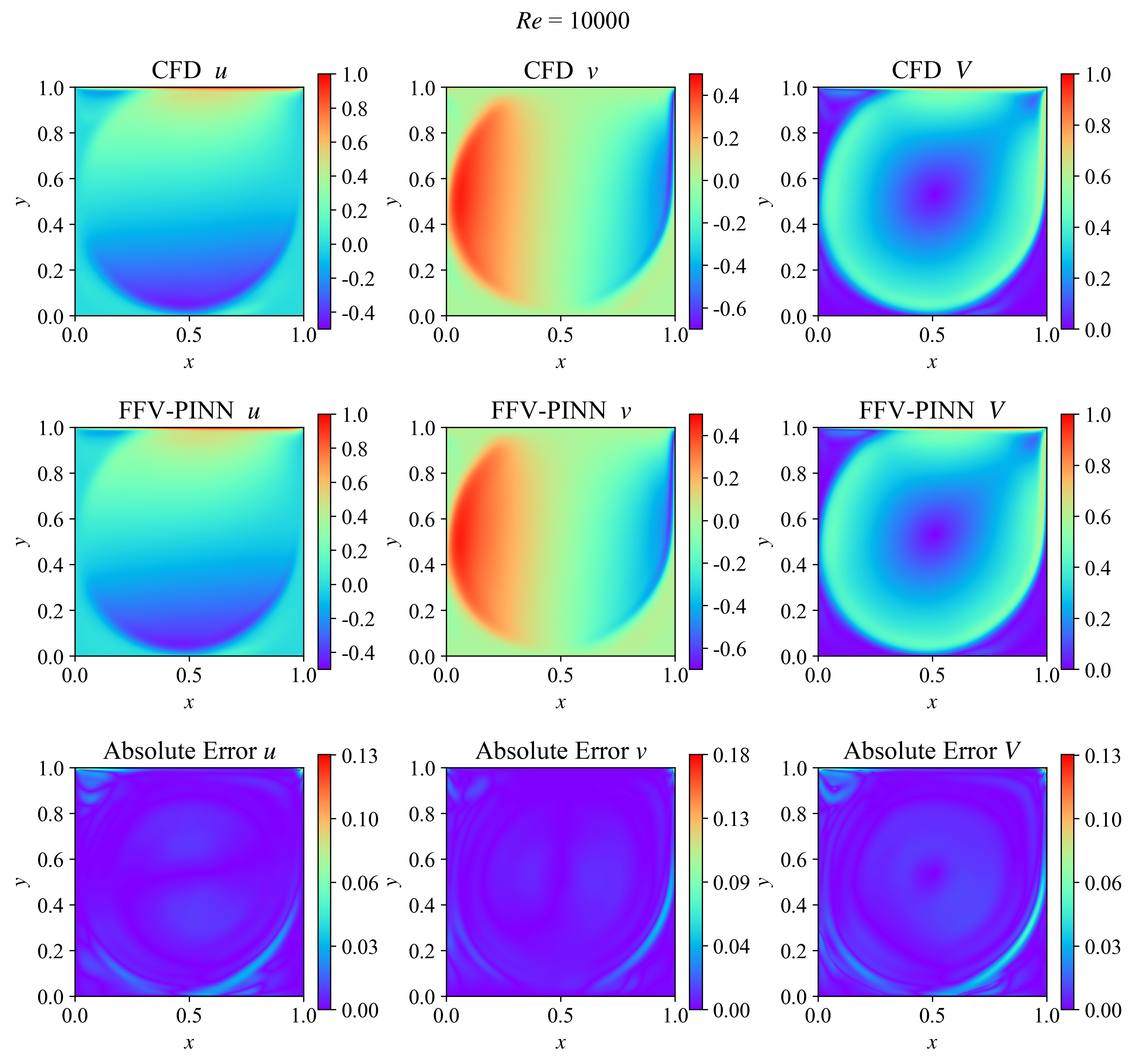}
    \caption{Lid-driven cavity flow at $Re$ = 10000. The left column represents the velocity component $u$, the middle column corresponds to the velocity component $v$, and the right column depicts the velocity magnitude $V$. Each row corresponds to, respectively, the solutions obtained via CFD, the proposed FFV-PINN method, and the absolute error between the two.}
    \label{fig:ldc_10000_contour}
\end{figure}

To further demonstrate the superior performance of FFV-PINN, we perform training for the lid-driven cavity flow problem at $Re$ = 7500 and $Re$ = 10000, as shown in Fig. \ref{fig:ldc_7500_contour} and \ref{fig:ldc_10000_contour}. In both cases, the $\lambda_{de }$, $\lambda_{bc}$, $\lambda_{rc}$, and $\alpha$ are set to 1.0, 1.0, 5.0, and 0.79, respectively. 
% It is noteworthy that the parameters employed in numerical experiments were determined empirically or through trial-and-error methods, rather than via a systematic optimization process. The primary objective of this study is to demonstrate the feasibility of the proposed framework, instead of achieving an optimal solution. 
To the best of our knowledge, this represents a groundbreaking achievement: the first successful application of PINNs to this benchmark problem at such high $Re$ numbers without incorporating any data. Particularly noteworthy is the case of $Re$ = 10000,  where the multiple vortexes with different scales are observed due to highly non-linearity, and also be a challenging problem even using CFD solver \citep{shankar_ldcreview2000l}. 
% This high $Re$ introduces substantial complexities for computational modeling, stemming from the inherent vortices and multi-scale nature. 

Fig. \ref{fig:ldc_7500_contour} shows contours for the lid-driven cavity flow at $Re$ = 7500. As shown in the figure, the velocity components ($u$ and $v$) and the velocity magnitude ($V$), obtained by the FFV-PINN, exhibit excellent agreement with those computed by high fidelity computational fluid dynamics (CFD) \citep{chiu2018CFD}. The contours of the velocity fields from both methods are virtually identical, indicating that the FFV-PINN method accurately captures the complex flow pattern at this high $Re$.  The consistently low absolute error across the computational domain further substantiates this quantitative agreement. Fig. \ref{fig:ldc_10000_contour} extends this analysis to $Re$ = 10000. Despite the increase in $Re$ and nonlinearity, FFV-PINN maintains consistency with the CFD benchmarks. Crucially, unlike prior works \citep{wang2023solution,jiang2023applications}, which augment PINNs with simulation data to stabilize training, our framework relies exclusively on boundary conditions—no additional training data is employed. This data-free predictive capacity underscores FFV-PINN's efficiency and distinguishes it from prior PINN work. This also fundamentally expands the potential for FFV-PINN to be applied across a wide array of fluid dynamics applications, especially in problems where the availability of high-quality experimental or simulation data is limited or prohibitive.

\begin{figure}[!ht]
    \centering
    \includegraphics[width=0.9\linewidth]{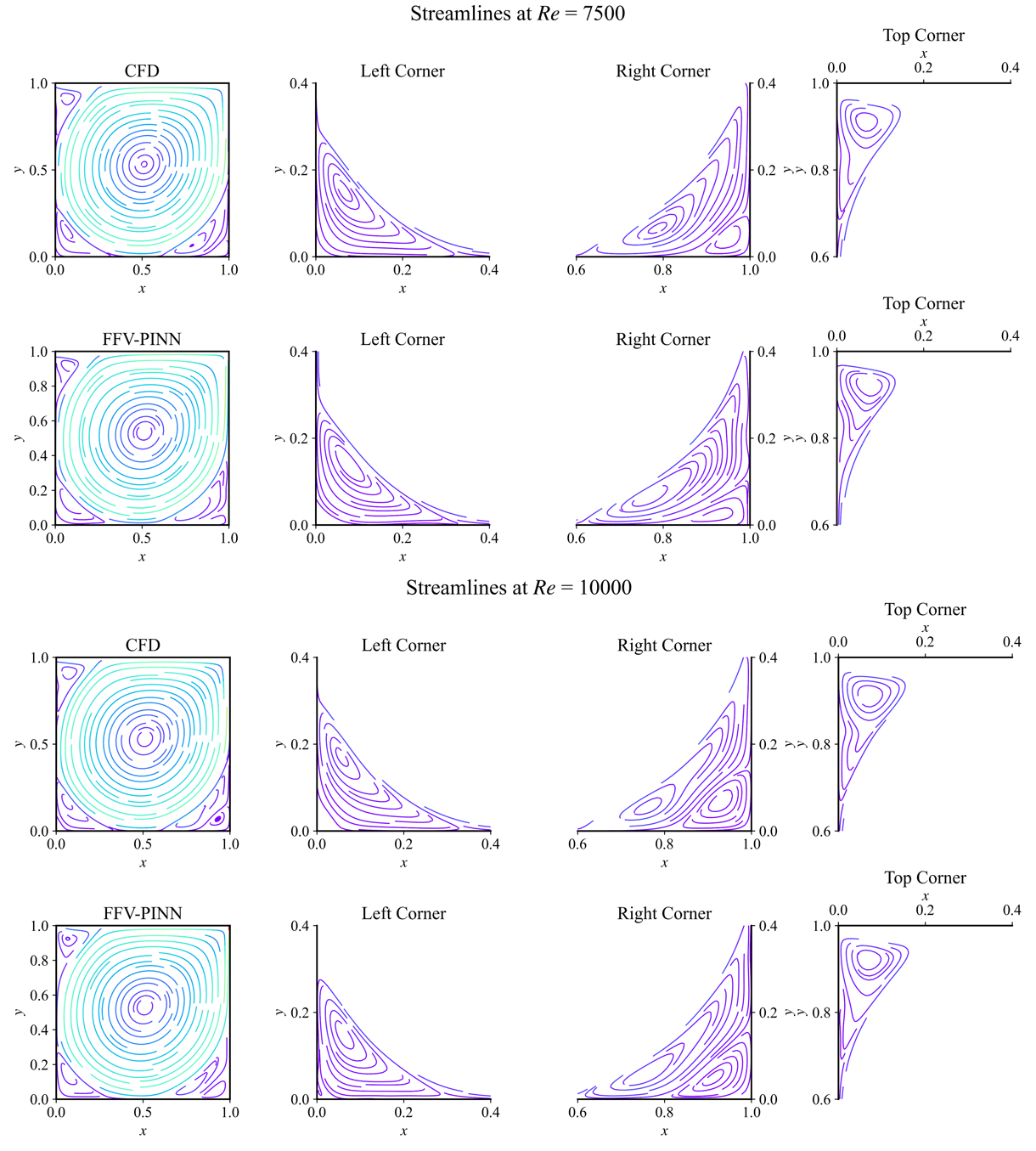}
    \caption{Streamlines comparison at $Re$ = 7500 and $Re$ = 10000. The first row shows the CFD results; the second row shows the results obtained by the FFV-PINN model. From left to right: global streamlines, and detailed streamlines of the bottom-left corner, bottom-right corner, and top-left corner, respectively.}
    \label{fig: streamline}
\end{figure}

Fig. \ref{fig: streamline} displays the streamlines for the lid-driven cavity flow at $Re$ = 7500 and $Re$ = 10000 respectively.
FFV-PINN demonstrates a remarkable ability to replicate the key flow structures that are observed in CFD results.
At both $Re$, FFV-PINN successfully captures the formation of the central primary vortex, which is a dominant feature of the flow pattern within the cavity. Additionally, the model accurately reproduces the secondary corner vortices that develop as a result of the interactions between the fluid and the stationary walls. 

\begin{figure}[!ht]
    \centering
    \includegraphics[width=1.0\linewidth]{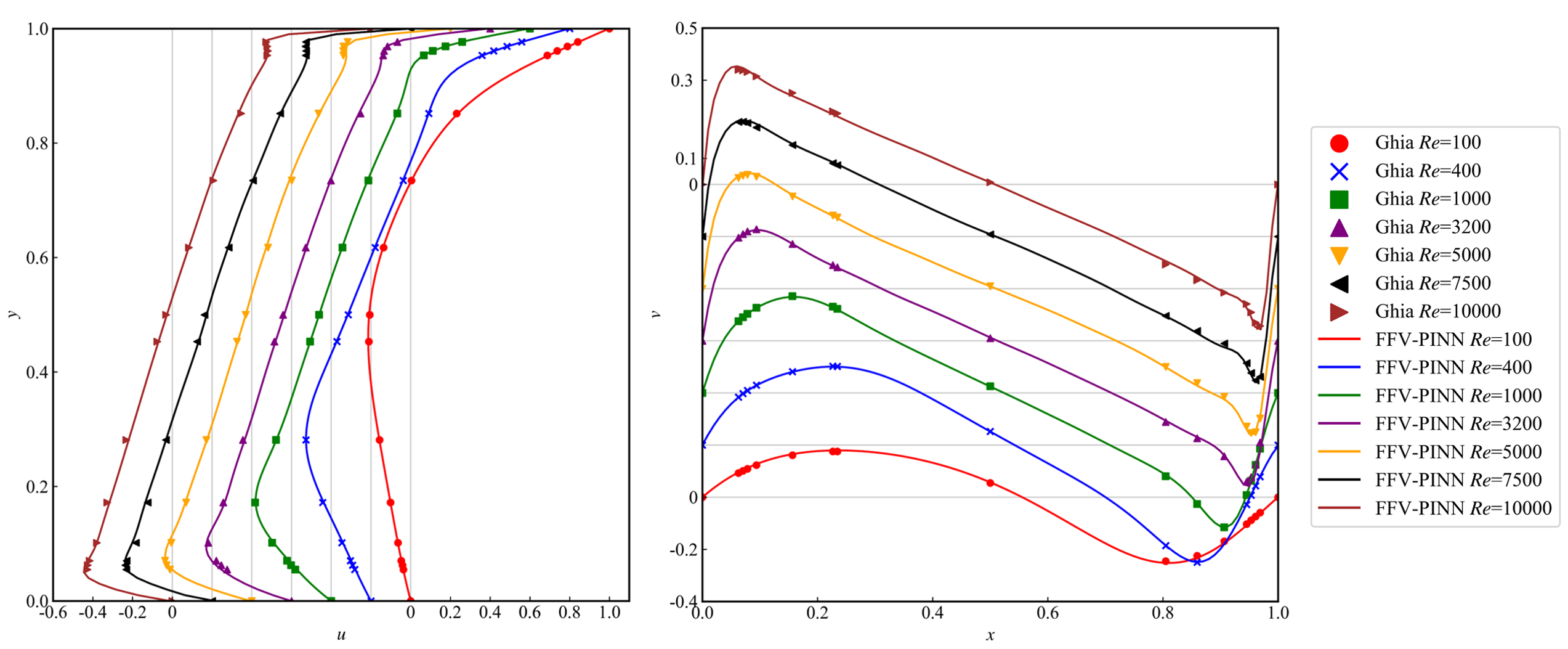}
    \caption{Comparison of the velocity profiles between the FFV-PINN and the numerical results from Ghia. Left: Comparison of $u$-velocity along vertical lines through the geometric center. Right: Comparison of profiles of $v$-velocity along horizontal lines through the geometric center.}
    \label{fig:uv_lines}
\end{figure}

Fig. \ref{fig:uv_lines} presents the velocity profiles for 
$u$ along vertical lines and $v$ along horizontal lines, both passing through the geometric center of the cavity.  Across the range of $Re$ considered, the FFV-PINN method demonstrates consistency with established numerical solutions reported by Ghia \citep{ghia1982high}. Specifically, the extreme --- both maximum and minimum values --- of the velocity components predicted by the FFV-PINN exhibits close quantitative correspondence to benchmark data.

\begin{figure}[!ht]
    \centering
    \includegraphics[width=1.0\linewidth]{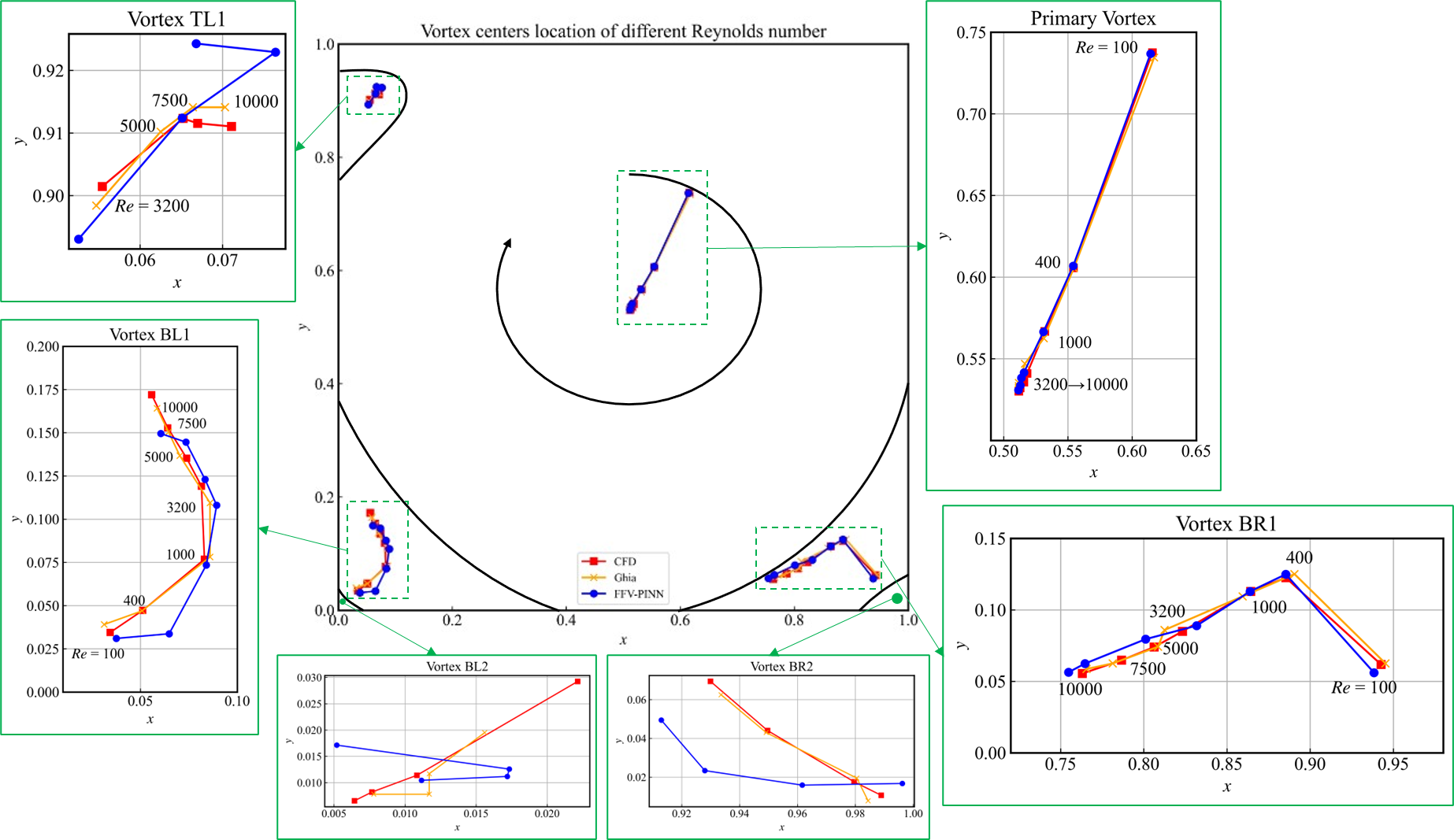}
    \caption{A comparative analysis of the vortex center locations across varying $Re$. The central plot displays the overall domain of the lid-driven cavity, with six enlarged subplots embedded within it. Each subplot focuses on a specific region of interest where secondary vortices are expected to form: the Top-Left corner (TL1), the Bottom-Left corner (BL1 and BL2), the Bottom-Right corner (BR1 and BR2), and the Primary Vortex core. Each subplot compares the vortex center locations obtained from CFD (red squares), the proposed FFV-PINN model (blue circles), and the Ghia (orange cross).   }
    \label{fig: vortex}
\end{figure}

Fig. \ref{fig: vortex} presents a detailed analysis of the evolution of vortex center positions, including primary vortex and secondary vortices, across multiple $Re$ ($Re$ = 100, 400, 1000, 3200, 5000, 7500, and 10000). The FFV-PINN method exhibits remarkable accuracy in predicting the vortex centers of the primary vortex, BL1, and BR1, with their displacement across $Re$ aligning closely with both CFD simulations and the benchmark data from Ghia. Discrepancies are noted for the TL1, BL2, and BR2 vortices, likely due to their relatively small magnitudes, which complicates the precise resolution of their positions. Despite these minor deviations, the overall consistency in vortex position underscores the effectiveness of the FFV-PINN method in capturing complex vortex structures across diverse flow conditions. Nonetheless, these discrepancies highlight potential avenues for further refinement, particularly in resolving fine-scale flow structures and enhancing the accuracy of secondary vortices. Addressing these aspects may further improve the model's precision and extend its applicability to even more complex fluid dynamics problems.
    
Table \ref{tab: LDC_error_table} provides a quantitative evaluation of the model's accuracy relative to ground truth CFD solutions \citep{chiu2018CFD}. These metrics are computed for the velocity components $u$, $v$, and the velocity magnitude $V$ across a range of $Re$.  In addition, the table details the training time and sample points used for each case. As expected, both MSE and relative $L_2$ error exhibit an increasing trend with $Re$, reflecting the greater challenge of accurately predicting increasingly complex flow behavior at higher $Re$. 
\begin{table}[!ht]
\centering
\caption{Quantitative evaluation of model accuracy for lid-driven cavity flow at various $Re$.  The sample points indicate the product of grid points. For 2D steady cases, this corresponds to the product of the number of grid points in the $x$ and $y$ directions.}
\label{tab: LDC_error_table}
% \vspace{0.5em}
\resizebox{\textwidth}{!}{
\begin{tabular}{ccccccccc}
\toprule
\multirow{2.5}{*}{$Re$} & \multicolumn{3}{c}{MSE} & \multicolumn{3}{c}{Relative $L_2$ Error} & \multirow{2.5}{*}{Time (s)} & \multirow{2.5}{*}{Sample points} \\
\cmidrule(lr){2-4} \cmidrule(lr){5-7} 
 & $u$ & $v$ & $V$ & $u$ & $v$ & $V$ & & \\
\midrule
100 & 4.77e-05 & 1.99e-05 & 2.80e-05 & 3.51e-02 & 2.88e-02 & 2.12e-02 & 129 & $51\times51$ \\
400 & 9.41e-06 & 7.22e-06 & 9.79e-06 & 1.49e-02 & 1.42e-02 & 1.12e-02 & 209 & $101\times101$ \\
1000 & 8.95e-06 & 7.61e-06 & 1.48e-05 & 1.36e-02 & 1.40e-02 & 1.30e-02 & 180 & $128\times128$ \\
3200 & 7.11e-05 & 6.28e-05 & 1.30e-04 & 3.78e-02 & 3.72e-02 & 3.69e-02 & 229 & $256\times256$ \\
5000 & 7.60e-05 & 7.00e-05 & 1.36e-04 & 4.05e-02 & 4.05e-02 & 3.90e-02 & 237 & $256\times256$ \\
7500 & 5.45e-05 & 4.51e-05 & 9.04e-05 & 3.44e-02 & 3.12e-02 & 3.13e-02 & 241 & $256\times256$ \\
10000 & 4.45e-05 & 4.43e-05 & 8.06e-05 & 3.19e-02 & 3.17e-02 & 3.03e-02 & 680 & $256\times256$ \\
\bottomrule
\end{tabular}
}
\end{table}
Despite this trend, the model maintains an excellent level of accuracy across the entire range of $Re$. At lower $Re$ (e.g., $Re$ = 100), the model exhibits very small errors, with MSE of $V$ = 2.80e-05 and relative $L_2$ error of $V$ = 2.12e-02, which is indicative of its ability to accurately capture laminar flow dynamics. Even at higher $Re$ (e.g., $Re$ = 10000),  where the flow becomes increasingly complex and turbulent, the model provides reasonably accurate predictions, with MSE of $V$ = 8.06e-05 and relative $L_2$ error of $V$ = 3.03e-02. 

An increase in training time with $Re$ is observed, which aligns with the greater computational demands required to resolve the finer flow details present at higher $Re$. This is also consistent with the increase in mesh resolution required for traditional CFD at higher $Re$ to accurately capture the increasingly complex flow structures. 
However, despite these challenges, the FFV-PINN method converges rapidly to an accurate solution across all $Re$. Even at $Re$=10000, the required computation time is only about 10 minutes.

To further demonstrate the capabilities and robustness of the FFV-PINN framework, the 2D unsteady lid-driven cavity flow at $Re$ = 400 and the 3D steady lid-driven cavity flow at $Re$ = 1000 are investigated. The governing equations for the 2D unsteady lid-driven cavity are expressed by:
\small
\begin{subequations} \label{eq:2dldc_unsteady}
    \begin{align}
        \frac{\partial u}{\partial x} + \frac{\partial v}{\partial y} &= 0 \label{eq:2dldc_unsteady-div} \\ 
         \frac{\partial u}{\partial t} +\frac{\partial \left(uu\right)}{\partial x} +  \frac{\partial \left(vu\right)}{\partial y} &= \frac{1}{Re}\left( \frac{\partial ^2u}{\partial x^2} + \frac{\partial ^2u}{\partial y^2} \right) - \frac{\partial p}{\partial x} \label{eq:2dldc_unsteady-m1} \\ 
        \frac{\partial  v}{\partial t} + \frac{\partial \left( u v\right)}{\partial x} +  \frac{\partial \left(v v\right)}{\partial y} &= \frac{1}{Re}\left( \frac{\partial ^2v}{\partial x^2} + \frac{\partial ^2v}{\partial y^2} \right) - \frac{\partial p}{\partial y}  \label{eq:2dldc_unsteady-m2} 
    \end{align}
\end{subequations}
Unlike the steady incompressible N-S equations, the unsteady form includes the temporal derivative term in the momentum equations, which is computed using AD.

\begin{figure}[!ht]
    \centering
    \includegraphics[width=1.0\linewidth]{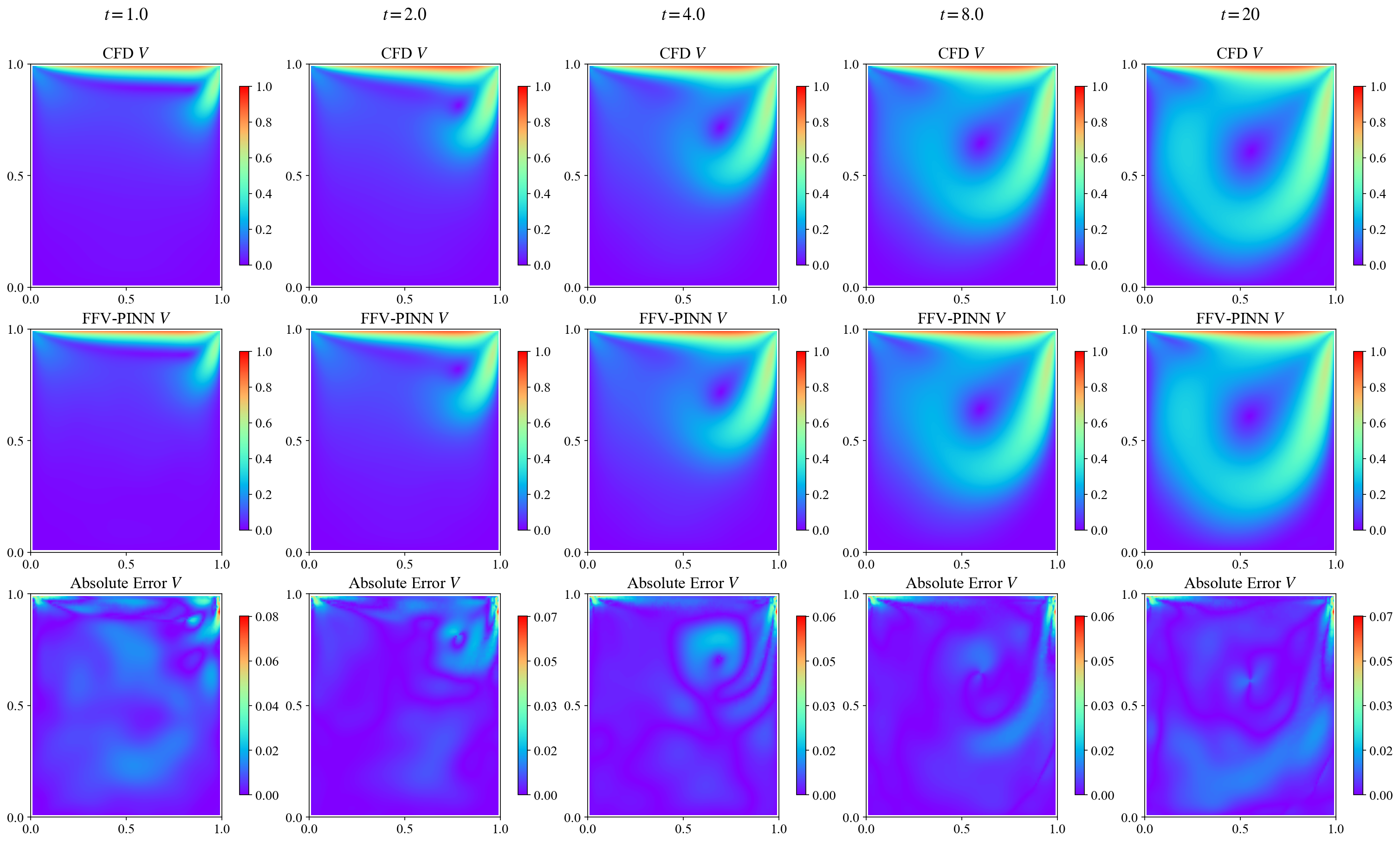}
    \caption{Comparison of the vertical magnitude ($V$) contours at different time instances ($t=1.0, 2.0, 4.0, 8.0, 20.0$) for the 2D unsteady lid-driven cavity flow ($Re=400$). The top row shows CFD results, the middle row shows FFV-PINN predictions, and the bottom row displays the absolute error between them.}
    \label{fig:ldc_unsteady_contour}
\end{figure}

Fig. \ref{fig:ldc_unsteady_contour} presents the contours of velocity magnitude $V$  at several representative time instances for the 2D unsteady lid-driven cavity flow at $Re$ = 400. The velocity fields predicted by the proposed FFV-PINN show excellent agreement with those obtained from high-fidelity CFD simulations \citep{chiu2018CFD} across all time steps. As time progresses, the FFV-PINN accurately captures the formation and evolution of the primary vortex, a hallmark feature of unsteady lid-driven cavity flows. Moreover, the absolute error distributions indicate that the FFV-PINN maintains uniformly low prediction errors across the spatial domain. Crucially, these errors remain stable over time, showing no evidence of temporal accumulation. This demonstrates the model’s ability to preserve accuracy over long-duration unsteady simulations, which is essential for modeling time-dependent fluid dynamics.

\begin{figure}[!ht]
    \centering
    \includegraphics[width=0.5\linewidth]{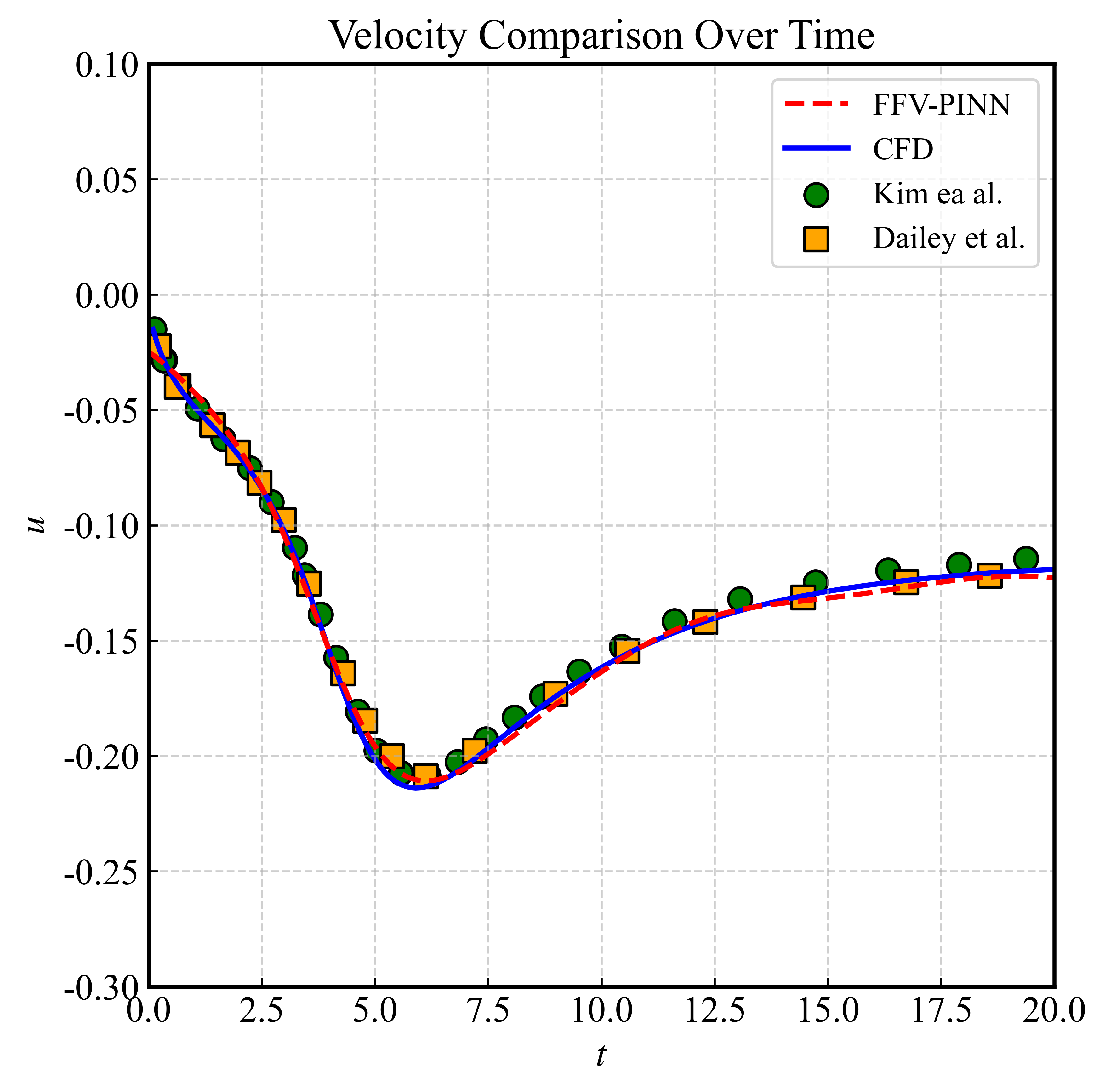}
\caption{Temporal evolution of the horizontal velocity component $u$ at the geometric center point $(0.5, 0.5)$ of the 2D cavity. The plot compares FFV-PINN predictions (red dashed line) with high-fidelity CFD results (blue solid line) and established benchmark data from Kim et al. \citep{kim2002implicit} (green circles) and Dailey et al. \citep{dailey1996evaluation} (orange squares).}
    \label{fig:unsteady_u}
\end{figure}

Fig~\ref{fig:unsteady_u} illustrates the temporal evolution of the horizontal velocity component $u$ at the geometric center of the cavity. The predictions from FFV-PINN closely match the CFD results\citep{chiu2018CFD} over the entire simulation period, and both show excellent agreement with benchmark data from Kim et al.\citep{kim2002implicit} and Dailey et al.\citep{dailey1996evaluation}. This consistency validates the accuracy and reliability of the FFV-PINN framework in capturing transient flow dynamics. 

\begin{figure}[!ht]
    \centering
    \includegraphics[width=0.5\linewidth]{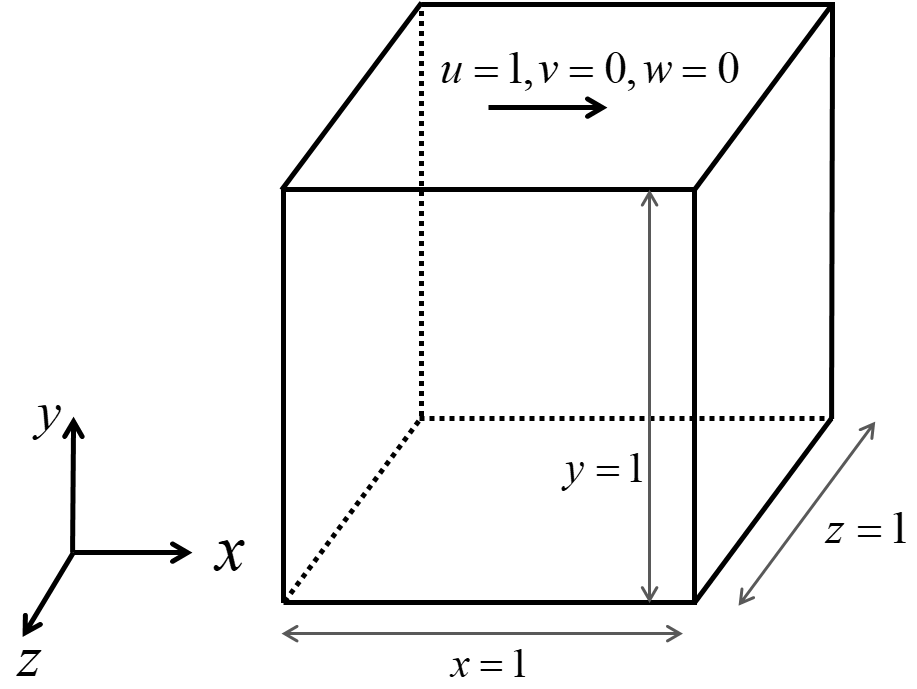}
\caption{Geometry of the 3D lid-driven cavity}
    \label{fig:3d_geo}
\end{figure}

The 3D lid-driven cavity problem is defined in a unit cubic domain \(\Omega = [0,1]^3\), where all six boundaries are solid walls, as illustrated in Fig.~\ref{fig:3d_geo}. The top lid, located at \(y = 1\), moves uniformly in the \(x\)-direction with a constant velocity \(u = 1\), while all other walls remain stationary and satisfy no-slip boundary conditions. The governing equations for this steady incompressible flow problem are given by:

\small
\begin{subequations} \label{eq:3dldc}
    \begin{align}
        \frac{\partial u}{\partial x} + \frac{\partial v}{\partial y} + \frac{\partial w}{\partial z} &= 0 \label{eq:3dldc_div} \\ 
        \frac{\partial (u u)}{\partial x} + \frac{\partial (v u)}{\partial y} + \frac{\partial (w u)}{\partial z} &= -\frac{\partial p}{\partial x} + \frac{1}{Re} \left( \frac{\partial^2 u}{\partial x^2} + \frac{\partial^2 u}{\partial y^2} + \frac{\partial^2 u}{\partial z^2} \right) \label{eq:3dldc-m1} \\
        \frac{\partial (u v)}{\partial x} + \frac{\partial (v v)}{\partial y} + \frac{\partial (w v)}{\partial z} &= -\frac{\partial p}{\partial y} + \frac{1}{Re} \left( \frac{\partial^2 v}{\partial x^2} + \frac{\partial^2 v}{\partial y^2} + \frac{\partial^2 v}{\partial z^2} \right) \label{eq:3dldc-m2}\\
        \frac{\partial (u w)}{\partial x} + \frac{\partial (v w)}{\partial y} + \frac{\partial (w w)}{\partial z} &= -\frac{\partial p}{\partial z} + \frac{1}{Re} \left( \frac{\partial^2 w}{\partial x^2} + \frac{\partial^2 w}{\partial y^2} + \frac{\partial^2 w}{\partial z^2} \right) \label{eq:3dldc-m3}
    \end{align}
\end{subequations}

\begin{figure}[!ht]
    \centering
    \includegraphics[width=1.0\linewidth]{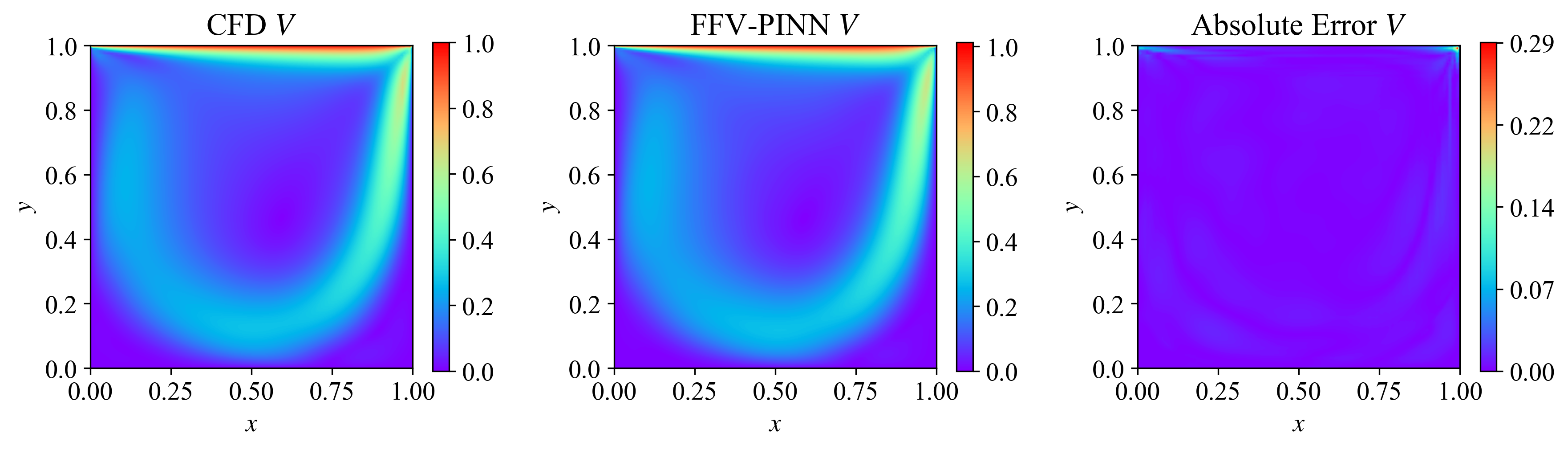}
\caption{Comparison of the velocity magnitude $V$ distributions on the $z=0.5$ plane obtained from CFD simulations, the  FFV-PINN model, and their corresponding pointwise absolute error}
    \label{fig:3d_V}
\end{figure}

Fig.~\ref{fig:3d_V} presents the contours of velocity magnitude $V$ on the $z$=0.5 plane for the 3D lid-driven cavity flow at $Re$ = 1000. The velocity fields predicted by the FFV-PINN show a high degree of consistency with CFD results \citep{chiu2018CFD}. This agreement suggests that the proposed approach is capable of maintaining reliable predictive performance in 3D scenarios, indicating its potential applicability to complex spatial problems. The pointwise absolute error distribution further supports this observation, showing uniformly low errors across the domain, with only a slight increase near the top-right corner. 

\begin{figure}[!ht]
    \centering
    \includegraphics[width=0.8\linewidth]{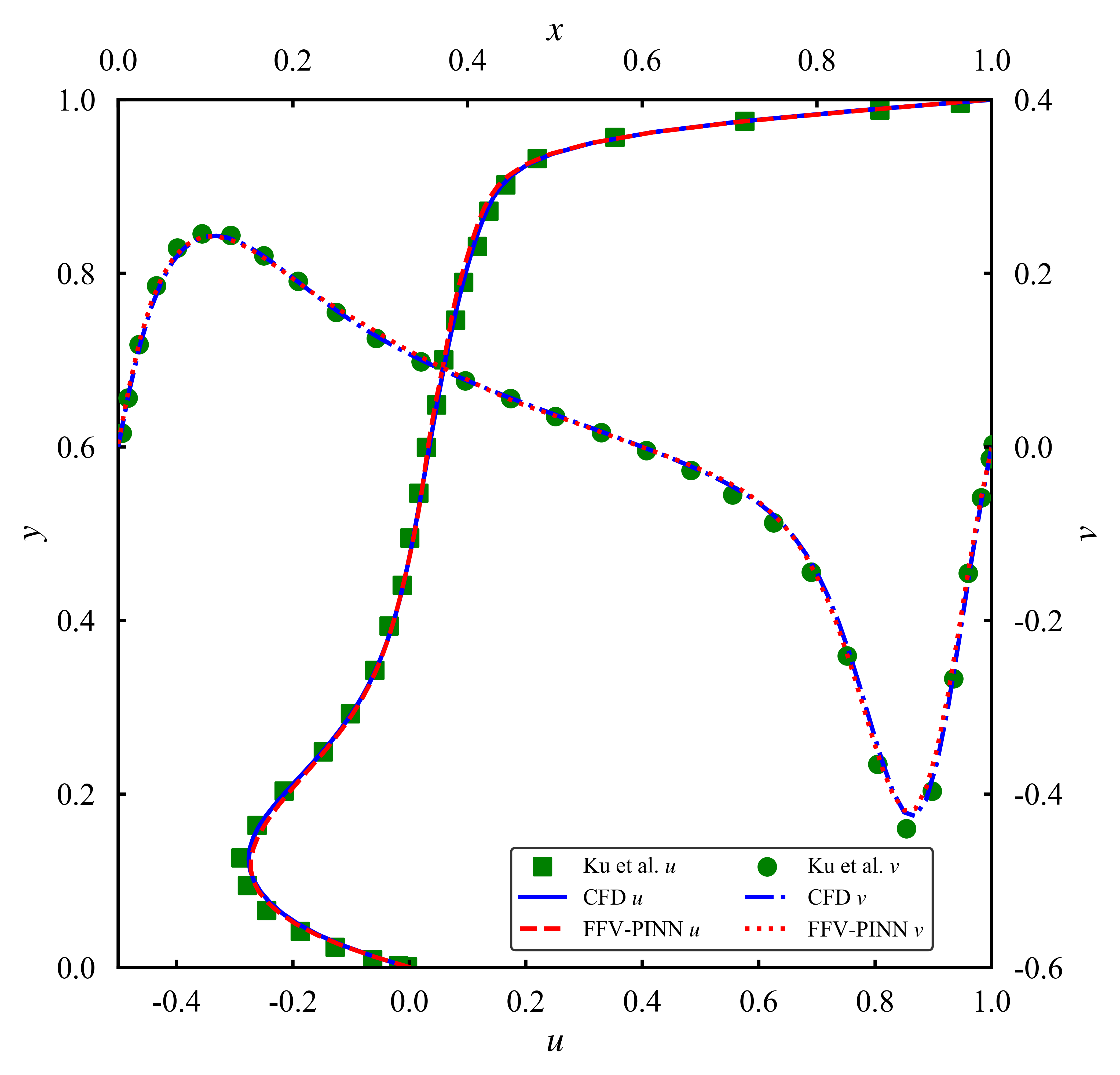}
\caption{
    Comparison of velocity profiles along the centerlines of a 3D lid-driven cavity flow at $Re = 1000$. The horizontal velocity component $u$ is plotted along the vertical centerline ($x=0.5$, $z=0.5$), with green squares representing the data from Ku et al. \citep{ku1987pseudospectral}, the blue solid line indicating CFD results, and the red dashed line showing predictions by the FFV-PINN. The vertical velocity component $v$ is plotted along the horizontal centerline  ($y=0.5$, $z=0.5$), with green circles denoting the data from Ku et al. \citep{ku1987pseudospectral}, the blue dash-dotted line indicating CFD results, and the red dash-dotted line representing FFV-PINN predictions.
}    
    \label{fig:3d_uv}
\end{figure}

Fig.~\ref{fig:3d_uv} presents a comparison of centerline velocity profiles predicted by the FFV-PINN with our in-house CFD results \citep{chiu2018CFD} and reference benchmark data from literature \citep{ku1987pseudospectral}. It is clearly evident from the figure that the predictions of the FFV-PINN show a high degree of consistency with those of Ku et al.\citep{ku1987pseudospectral} and the CFD for both the horizontal velocity component $u$ and the vertical component $v$, highlighting the remarkable accuracy and robustness of the FFV-PINN model in capturing the complex flow dynamics.

Table~\ref{tab:2d_unsteady_3d_results} presents a quantitative evaluation of the FFV-PINN on two benchmark problems: the 2D unsteady and 3D steady lid-driven cavity flow. Evaluation metrics include the relative $L_2$ error and MSE for the velocity components $u$, $v$, and $w$, as well as the velocity magnitude $V$. Training time and sample points are also reported.

In terms of relative $L_2$ error, the FFV-PINN demonstrates good accuracy in both the 2D unsteady and 3D steady lid-driven cavity flow. Specifically, the relative $L_2$ error for the velocity magnitude $V$ is $3.12\times10^{-2}$ in the 2D unsteady case and $5.09\times10^{-2}$ in the 3D steady case. Although the error in the 3D case is slightly higher, this is expected given the increased complexity associated with higher spatial dimensions and the more intricate flow structures present in the 3D problem. From the perspective of MSE, a similar trend is observed. The MSE for the velocity magnitude $V$ increases from $5.60\times10^{-5}$ in the 2D case to $1.23\times10^{-4}$ in the 3D case. Nevertheless, the error remains within an acceptable range, indicating that the model maintains accuracy despite the increased spatial complexity.

Furthermore, the training time remains relatively low for both cases, with the 3D case requiring 506s and the 2D case 564s, highlighting the computational efficiency of the FFV-PINN. In summary, the proposed model not only achieves high numerical accuracy in both the 2D unsteady and the more complex 3D steady lid-driven cavity flow but also maintains consistently low training times across different scenarios. These results underscore the model’s strong generalization capability across various flow regimes, as well as its robustness and potential for a wide range of applications in flow prediction tasks spanning both steady and transient 2D and 3D flows.

\begin{table}[!ht]
    \centering
    \caption{Quantitative evaluation of model accuracy for 2D unsteady and  3D steady lid-driven cavity flow. The sample points indicate the product of grid points. For 2D unsteady cases, this corresponds to the product of the number of grid points in time ($t$) and the spatial dimensions ($x$ and $y$), while for 3D cases, it denotes the product of the grid points in the $x$, $y$, and $z$ directions.}
    \renewcommand{\arraystretch}{1.2}
    \label{tab:2d_unsteady_3d_results}
    \large
    \resizebox{\textwidth}{!}{
    \begin{tabular}{cccccccccc}
        \toprule
         Case & \multicolumn{4}{c}{Relative $L_2$ Error} & \multicolumn{4}{c}{MSE} & \multirow{2.2}{*}{Time}\\
        \cmidrule(lr){2-5} \cmidrule(lr){6-9}
        (Sample points)& $u$ & $v$ & $w$ & $V$ & $u$ & $v$ & $w$ & $V$ &  \\
        \midrule
        2D unsteady LDC & \multirow{2.2}{*}{3.74e-02} & \multirow{2.2}{*}{3.74e-02} & \multirow{2.2}{*}{-} & \multirow{2.2}{*}{3.12e-02}& \multirow{2.2}{*}{4.65e-05} &\multirow{2.2}{*}{3.39e-05}  & \multirow{2.2}{*}{-} &\multirow{2.2}{*}{5.60e-05}  & \multirow{2.2}{*}{564s}  \\
       $\left(200\times100\times100\right)$ &   &   &   &   &   &    &   &   &   \\
        3D steady LDC & \multirow{2.2}{*}{5.52e-02} & \multirow{2.2}{*}{5.94e-02 }&\multirow{2.2}{*}{2.25e-01}  & \multirow{2.2}{*}{5.09e-02} & \multirow{2.2}{*}{8.78e-05} & \multirow{2.2}{*}{6.49e-05 }&\multirow{2.2}{*}{1.60e-05}  &\multirow{2.2}{*}{1.23e-04}  & \multirow{2.2}{*}{506s}  \\
        $\left(128\times128\times128\right)$  &   &   &   &   &   &   &   &   &   \\        
        \bottomrule
    \end{tabular}
    }
\end{table}

\subsection{Backward-facing step flow} 
\begin{figure}[!ht]
    \centering
    \includegraphics[width=0.85\linewidth]{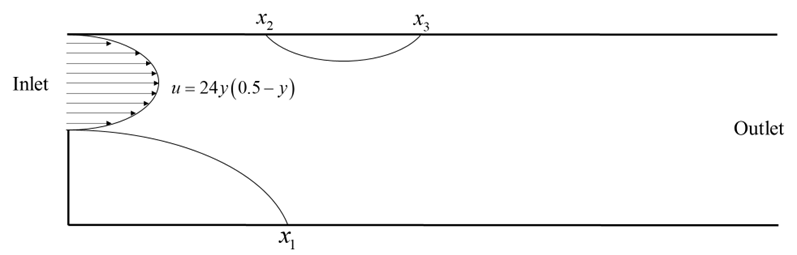}
    \caption{Schematic representation of the backward-facing step flow problem. A fully developed parabolic velocity profile, defined by $u = 24y(0.5 - y)$, is imposed at the inlet. Key locations of interest are denoted as $x_1$, $x_2$, and $x_3$, which typically represent the detachment and reattachment points.}
    \label{fig:bfs}
\end{figure}

The backward-facing step flow serves as another critical benchmark for evaluating the capability of numerical methods to resolve separated flows. In this study, we consider flow within a $20 \times 1$ unit channel, with a fully developed parabolic velocity profile imposed at the inlet. The problem is schematically illustrated in Fig. \ref{fig:bfs}. As the fluid flows over the backward-facing step, the abrupt expansion of the channel cross-section induces flow separation at the step corner (i.e., the point where the expansion begins). 

At a steady state, a primary recirculation zone immediately forms at the downstream of the step. Its extension is characterized by the location $x_1$.  With increasing $Re$, flow bifurcation occurs, leading to the development of a secondary vortex structure along the upper channel wall between $x_2$ and $x_3$.  Despite having different boundary conditions, the backward-facing step flow shares the same physical laws as the lid-driven cavity flow, both being governed by the steady-state, 2D, incompressible N-S equations, as detailed in Eq.\ref{eq:2dns}.

Table \ref{tab:backward_facing_step} presents comparative results of key flow characteristics in the backward-facing step problem, specifically focusing on the detachment and reattachment points ($x_1$, $x_2$, and $x_3$) obtained from different PINN models. The predicted $x_1$ positions from our FFV-PINN show remarkable consistency with CFD~\citep{chiu2018CFD} results at $Re$ = 200 and $Re$ = 400. Furthermore, the FFV-PINN's predicted $x_2$ and $x_3$ values at $Re$ = 400 also fall within the range reported by other CFD methods, further demonstrating the accuracy of our model in capturing flow separation and reattachment dynamics.
\begin{table}[ht]
\centering
\caption{Comparison of detachment and reattachment points from PINN-based predictions and benchmark solutions for the backward-facing step problem}
\label{tab:backward_facing_step}
% \resizebox{\textwidth}{!}{
\footnotesize
\begin{tabular}{cccccc}
\toprule
\multirow{2.5}{*}{References} & \multicolumn{1}{c}{$200$} & \multicolumn{3}{c}{$400$} \\
\cmidrule(lr){2-2} \cmidrule(lr){3-5} 
 & $x_1$ & $x_1$ & $x_2$ & $x_3$ \\
\midrule
CFD, Barton \citep{barton1995numerical} & 2.565 & 4.255 & 4.045 & 5.11  \\
CFD, Barber et al.\citep{barber2001numerical} &  2.635 & 4.265 & 4.065 & 5.07 \\
CFD, Erturk \citep{erturk2008numerical} &  2.491 & 4.119 & 3.866 & 5.019 \\
CFD, Chiu \citep{chiu2018CFD} &  2.608 & 4.305 & 4.091 & 5.122 \\
CAN-PINN \citep{chiu2022can}& 2.604 & 4.292 & 4.120 & 5.161 \\
LSFD-PINN \citep{xiao2024least}& 2.614 & 4.243 & 4.113 & 5.071 \\
% RBFDQ-PINN-I \citep{xiao2023RBFDQ}& 2.636 & 4.269 & 4.097 & 5.100 \\
RBFDQ-PINN-II \citep{xiao2023RBFDQ}&  2.632 & 4.212 & 4.069 & 5.130 \\
Ours &  2.609 & 4.307 & 4.145 & 5.089 \\
\bottomrule
\end{tabular}
% }
\end{table}

\begin{figure}[!ht]
    \centering
    \includegraphics[width=1.0\linewidth]{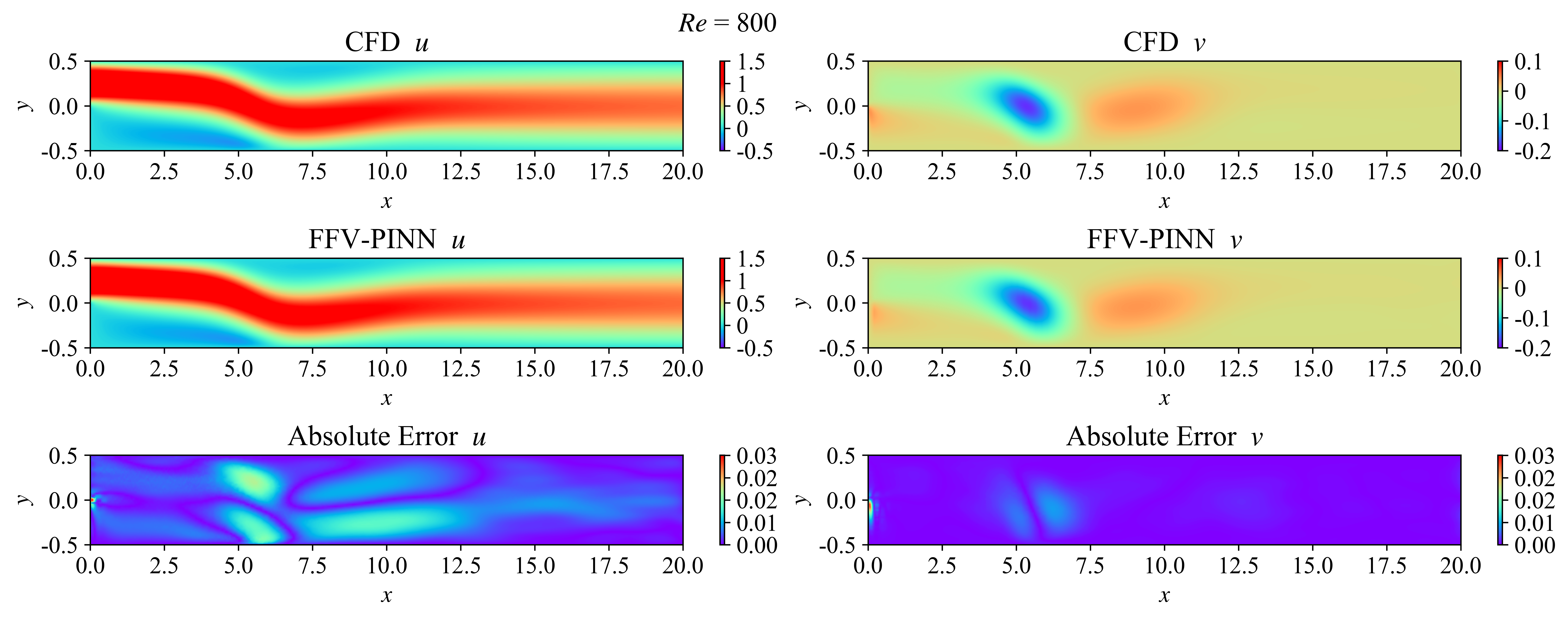}
    \caption{Velocity contour and absolute point-wise error distribution for the backward-facing step flow at $Re$ = 800. From top to bottom, the figures are as follows: $u$-velocity (left) and $v$-velocity (right) contours from CFD; $u$-velocity and $v$-velocity contours from the FFV-PINN; absolute point-wise error in $u$-velocity and $v$-velocity.}
    \label{fig:bfs_contour_800}
\end{figure}

Fig. \ref{fig:bfs_contour_800} presents the contours of the backward-facing step flow at $Re$ = 800. In this case, the $\lambda_{de }$, $\lambda_{bc}$, $\lambda_{rc}$, and $\alpha$ are set to 10.0, 10.0, 1.0, and 0.95, respectively. The FFV-PINN demonstrates a high level of accuracy in capturing both the $u$-velocity and $v$-velocity profile. The absolute error plots highlight the low discrepancies between FFV-PINN predictions and the CFD benchmarks, further validating the model's ability to capture overall flow behavior accurately. 

\begin{figure}[!ht]
    \centering
    \includegraphics[width=1.0\linewidth]{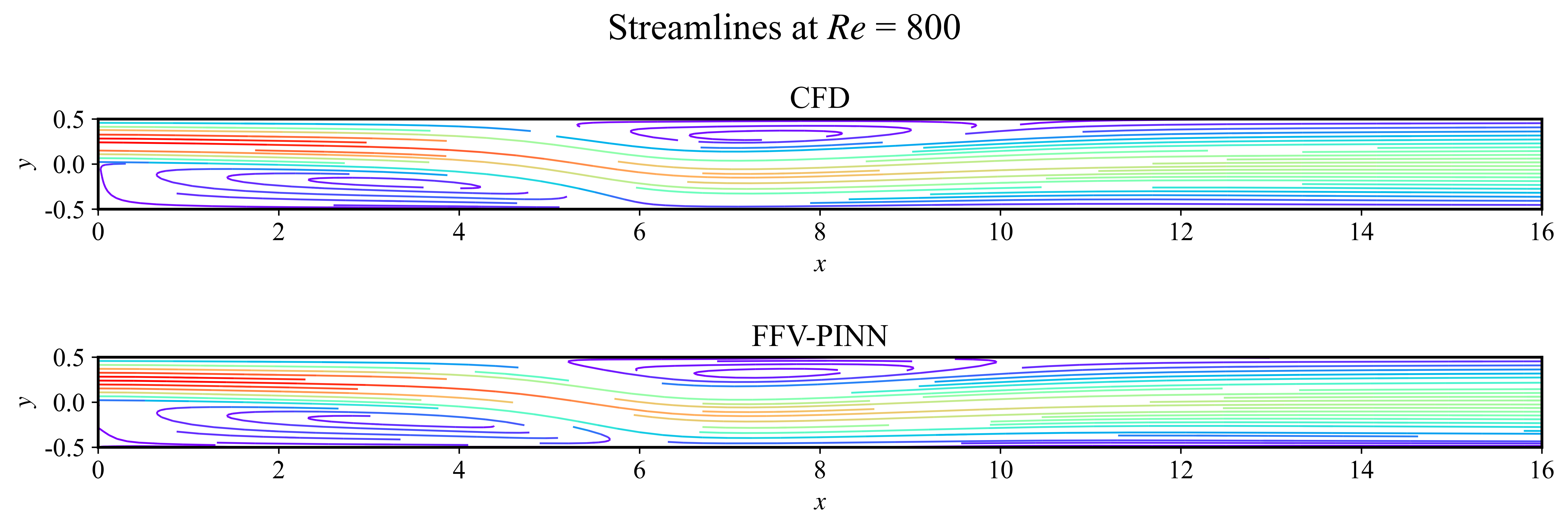}
    \caption{Streamlines for the backward-facing step flow at $Re$ = 800. The top panel presents the streamlines obtained from the CFD solution of Chiu \citep{chiu2018CFD}, while the bottom panel shows the streamlines from the FFV-PINN model. The color of the streamlines represents the velocity magnitude $V$.}
    \label{fig:bfs_streamlines_800}
\end{figure}

Fig. \ref{fig:bfs_streamlines_800} presents streamline visualizations illustrating the flow patterns associated with the backward-facing step problem at $Re$ = 800. The results obtained from the FFV-PINN exhibit a strong qualitative agreement with the predictions generated by CFD simulations. Significantly, the FFV-PINN model accurately replicates the formation of the recirculation zone immediately behind the step, as well as the emergence of a secondary eddy in the upper mid-region of the channel. These flow structures are fundamental features of backward-facing step flow, highlighting the model’s capability to capture the essential dynamics of separated flows with a high degree of fidelity.

\begin{figure}[!ht]
    \centering
    \includegraphics[width=1.0\linewidth]{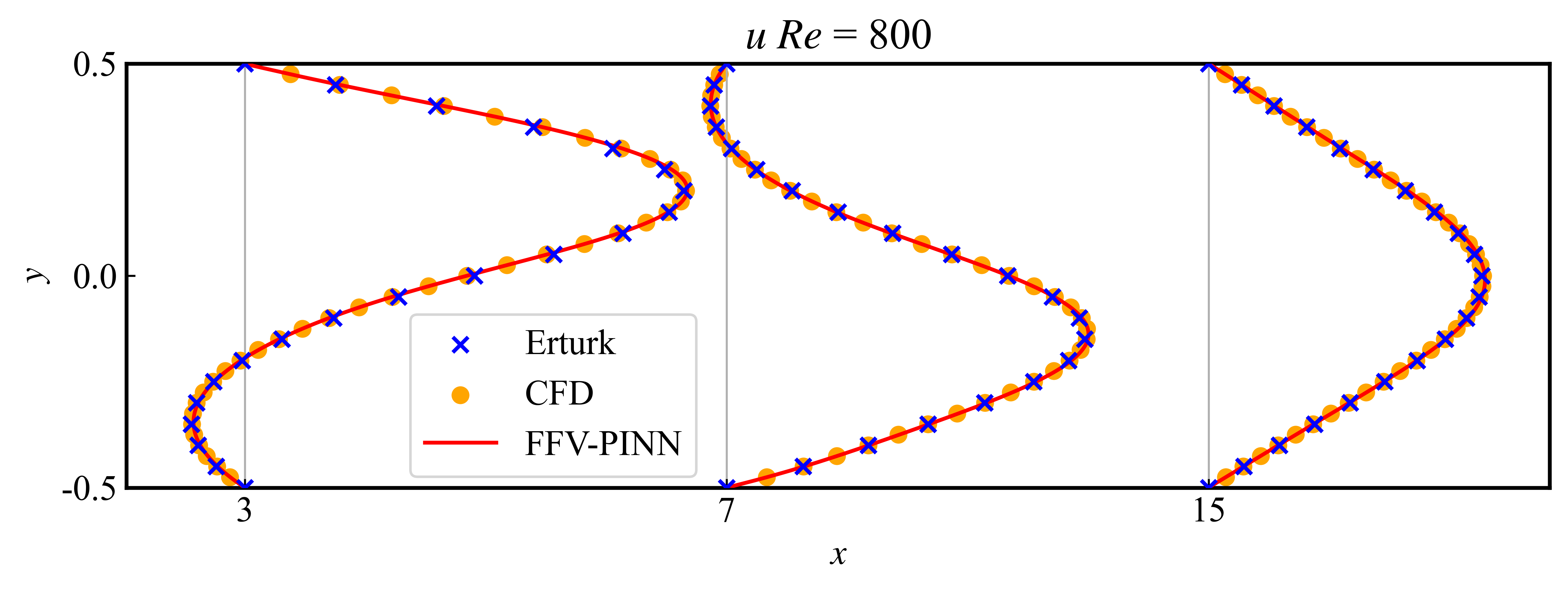}
    \caption{Comparison of $u$-velocity profiles at $Re$ = 800 for the backward-facing step flow.  Profiles are shown at various $x$-locations ($x$ = 3, $x$ = 7, and $x$ = 15) along the channel, comparing results from the FFV-PINN (red line), CFD solution of Chiu \citep{chiu2018CFD}(orange circles), and benchmark data from Erturk \citep{erturk2008numerical} (blue crosses).}
    \label{fig:bfs_u_800}
\end{figure}

Fig \ref{fig:bfs_u_800} displays a systematic comparison of the streamwise velocity $u$ profiles at $Re$ = 800, evaluated at three critical downstream positions ($x$ = 3, $x$ = 7, and $x$ = 15). At $x$ = 3, immediately downstream of the step, the $u$-velocity distribution exhibits a steep shear layer near the bottom wall, marking the formation of the primary recirculation vortex. This region is characterized by flow reversal and serves as the dominant separation zone. 
Further downstream at $x$ = 7, a secondary vortex emerges near the upper wall, as evidenced by localized velocity inflection points. At $x$ = 15, the flow achieves full redevelopment. The strong agreement observed across all $x$-locations  between the FFV-PINN results and both the CFD solution of Chiu \citep{chiu2018CFD} and the benchmark data from Erturk \citep{erturk2008numerical} highlights the model's ability to accurately predict the velocity distribution within the channel.  This consistency further validates the effectiveness of the FFV-PINN in resolving key flow characteristics, including shear layer development and recirculation dynamics, across different regions of the flow domain.

Table \ref{tab:bfs_error_table} presents a quantitative evaluation of the FFV-PINN model accuracy for backward-facing step flow at various $Re$ by utilizing CFD solutions as ground truth~\citep{chiu2018CFD} . The results indicate that the model maintains very low errors across all investigated $Re$. Specifically, for $Re = 100$, the relative $L_2$ error for the velocity components $u$ and $v$ are 1.63e-03 and 1.12e-02 respectively. Even at $Re = 800$, the errors remain small, with 8.41e-03  for $u$ and 3.48e-02  for $v$. The training time remains constant at 171 seconds for all cases, indicating that the model's efficiency does not degrade with increasing $Re$. Additionally, the error values show no significant fluctuations as $Re$ increases, demonstrating that the model maintains stable performance across different flow conditions. 
\begin{table}[ht]
\centering
\caption{Quantitative evaluation of model accuracy for backward-facing step flow at various $Re$ }
\label{tab:bfs_error_table}
% \vspace{0.5em}
\resizebox{\textwidth}{!}{
\begin{tabular}{ccccccccc}
\toprule
\multirow{2.5}{*}{$Re$} & \multicolumn{3}{c}{MSE} & \multicolumn{3}{c}{Relative $L_2$ Error} & \multirow{2.5}{*}{Time (s)} & \multirow{2.5}{*}{Sample points}\\
\cmidrule(lr){2-4} \cmidrule(lr){5-7} 
 & $u$ & $v$ & $V$ & $u$ & $v$ & $V$ & &  \\
\midrule
100 & 8.67e-07 & 1.02e-07 & 8.72e-07 & 1.63e-03 & 1.12e-02 & 1.64e-03 & 171 & 801$\times$41 \\
200 & 2.51e-06 & 1.64e-07 & 2.32e-06 & 2.72e-03 & 1.61e-02 & 2.61e-03 & 171 & 801$\times$41 \\
400 & 5.66e-06 & 3.78e-07 & 5.48e-06 & 3.93e-03 & 2.55e-02 & 3.87e-03 & 171 & 801$\times$41 \\
600 & 9.41e-06 & 3.37e-07 & 9.32e-06 & 4.89e-03 & 2.08e-02 & 4.86e-03 & 171 & 801$\times$41 \\
800 & 2.97e-05 & 1.25e-06 & 2.95e-05 & 8.41e-03 & 3.48e-02 & 8.38e-03 & 171 & 801$\times$41 \\
\bottomrule
\end{tabular}
}
\end{table}
\subsection{Natural convection in a unit square} 
\begin{figure}[!ht]
    \centering
    \includegraphics[width=0.5\linewidth]{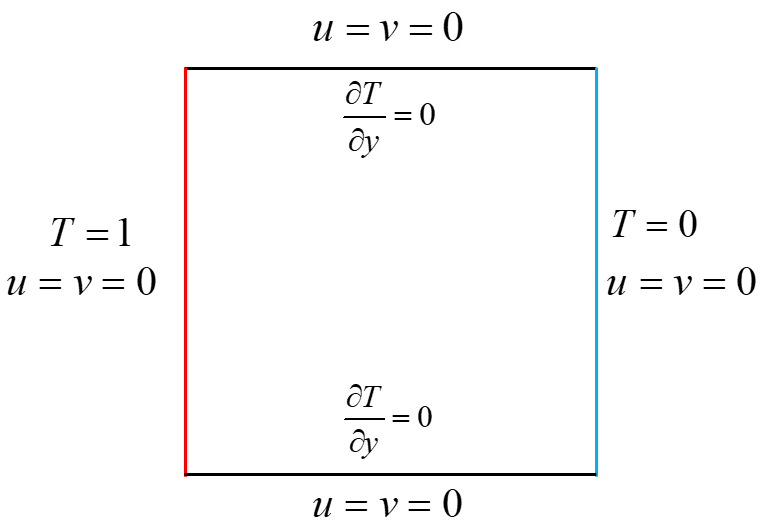}
    \caption{Schematic of the natural convection problem in a square cavity. The left and right walls are maintained at constant temperatures $T=1$ and $T=0$, respectively, while the top and bottom walls are adiabatic with $\frac{\partial T}{\partial y} =0$. No-slip boundary conditions ($u=v=0$) are imposed on all four walls.}
    \label{fig:nc}
\end{figure}

Natural convection is a vital, multi-physics phenomenon in both the natural sciences and engineering fields, involving the interplay between fluid flow and heat transfer. Addressing such problems requires solving a multi-physics scenario with thermal-fluid interactions as per the N-S equations and energy equation. For the current problem, the computational domain is a unit square, as shown in Fig \ref{fig:nc}. The vertical boundaries are maintained at distinct temperatures: the left wall is set to $T(x=0,y) = 1$ while the right wall remains at $T(x=1,y)=0$. The horizontal surfaces are characterized by zero thermal flux conditions $(\frac{\partial T}{\partial n}=0)$, creating thermally adiabatic interfaces that prevent heat exchange with the external environment. The governing equations for this system are expressed by:
\small
\begin{subequations} \label{eq:2dnc}
    \begin{align}
        &\frac{\partial u}{\partial x} + \frac{\partial v}{\partial y} = 0 \label{eq:2dnc-div} \\ 
        &u \frac{\partial u}{\partial x} + v \frac{\partial u}{\partial y} = \sqrt{\frac{Pr}{Ra}}\left( \frac{\partial ^2u}{\partial x^2} + \frac{\partial ^2u}{\partial y^2} \right) - \frac{\partial p}{\partial x} \label{eq:2dnc-m1} \\ 
        &u \frac{\partial v}{\partial x} + v \frac{\partial v}{\partial y} = \sqrt{\frac{Pr}{Ra}}\left( \frac{\partial ^2v}{\partial x^2} + \frac{\partial ^2v}{\partial y^2} \right) - \frac{\partial p}{\partial y} + T \label{eq:2dnc-m2} \\
        &u \frac{\partial T}{\partial x} + v \frac{\partial T}{\partial y} = \frac{1}{\sqrt{PrRa}}\left( \frac{\partial ^2T}{\partial x^2} + \frac{\partial ^2T}{\partial y^2} \right) \label{eq:2dnc-T}
    \end{align}
\end{subequations}
where $Pr$ denotes the Prandtl number representing the ratio of momentum diffusivity to thermal diffusivity and is fixed to 0.71. $Ra$ represents the Rayleigh number describing the relationship between buoyancy and viscosity of the fluid. An increase in $Ra$ signifies that the buoyancy forces, which drive the flow due to temperature differences, become significantly stronger relative to the viscous forces that resist motion. This transition can lead to a variety of nonlinear effects that can be quite complex and multifaceted.

\begin{figure}[!ht]
    \centering
    \includegraphics[width=1.0\linewidth]{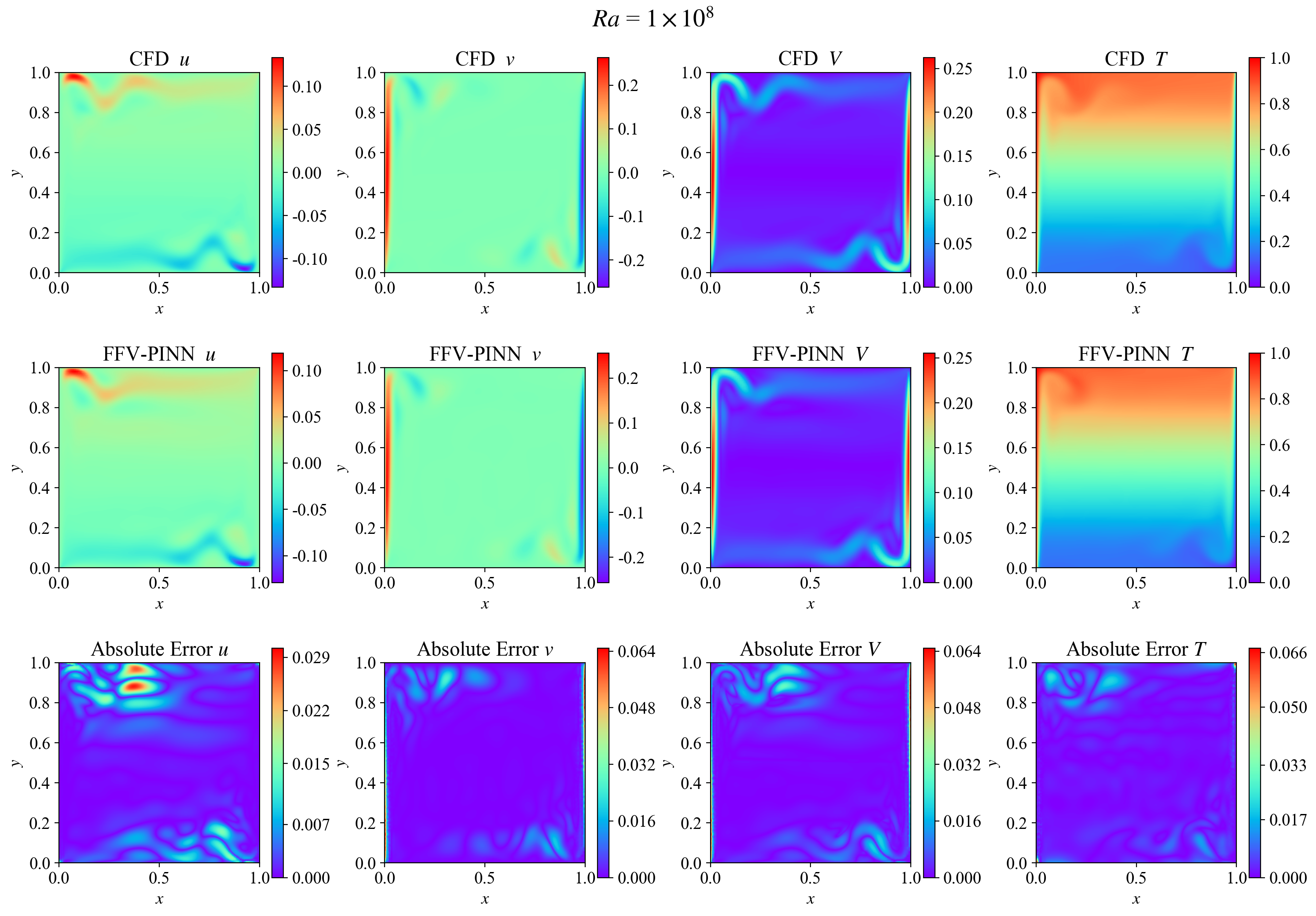}
    \caption{The velocity and temperature comparison for natural convection in a square cavity at $Ra = 10^8$.  The figure presents a comparative analysis of $u$-velocity, $v$-velocity, velocity magnitude $V$, and temperature $T$ contours obtained from CFD solution of Chiu\citep{chiu2018CFD} (top row), the FFV-PINN model (middle row), and the absolute error between them (bottom row).}
    \label{fig:nc_contour_1e8}
\end{figure}

Fig \ref{fig:nc_contour_1e8} provides contours for the evaluation of FFV-PINN in simulating natural convection phenomena at a high $Ra$ ($Ra = 10^8$). The $\lambda_{de }$, $\lambda_{bc}$, $\lambda_{rc}$, and $\alpha$ are set to 1.0, 1.0, 1.0, and 0.84, respectively. The FFV-PINN results demonstrate excellent agreement with CFD benchmarks~\citep{chiu2018CFD} in reconstructing the temperature distribution and velocity field. Notably, the FFV-PINN model effectively captures thin thermal boundary layers, which are critical for accurately simulating heat transfer in buoyancy-driven flows. In addition, the low error magnitudes also confirm the model's capacity to solve problems involving coupled flow and heat transfer.

Table \ref{tab:numerical_comparison} presents a comparison of the average Nusselt number (\(\overline{Nu}\)) and the maximum vertical velocity (\(v_{\max}\)) obtained from the proposed FFV-PINN model against established reference values~\citep{kalita2001fully} and CFD results~\citep{chiu2018CFD}. The FFV-PINN model exhibits excellent agreement with benchmark data for \(\overline{Nu}\), t. Specifically, at $Ra = 10^6$, FFV-PINN yields an $\overline{Nu}$  of 8.695, closely approximating the reference value of 8.829 and the CFD result of 8.785. Furthermore, by employing a refined \( 202 \times 202 \) grid at higher $Ra$ ($5 \times 10^7 $ and $ 10^8$), the model produces predictions of $\overline{Nu} = 24.87$ and $\overline{Nu} = 29.50$ respectively, which are very close to the CFD results of $\overline{Nu} = 25.27$  and $\overline{Nu} = 30.34$. These results underscore the accuracy of our FFV-PINN model in capturing the complex physics of natural convection, even under challenging high \( Ra \) conditions. 

For the \(v_{\max}\) along the horizontal midline, it is noted that due to the fact that the non-dimensional equations we derived are different from \citep{kalita2001fully,chiu2018CFD}, the scaling factor of $\sqrt{RaPr}$ is employed for comparisons. At $Ra$ = $10^7$, using a $75 \times75$ grid, the FFV-PINN model slightly underestimates $v_{max}$ compared to the CFD results. However, upon refining the grid resolution to $202 \times 202$, the FFV-PINN predictions become nearly identical to those from CFD. At higher $Ra$ ($5 \times 10^7 $ and $ 10^8$), the model continues to underestimate $v_{max}$ relative to the CFD outcomes. Despite these discrepancies, it is important to note that the magnitude of the differences is relatively small, and the errors remain within acceptable limits. In summary, these comparisons indicate that our method produces results that are very close to established benchmarks. The observed differences between numerical methods may be attributed to grid resolution and increased nonlinearity at higher $Ra$. A more detailed comparison is provided in \ref{sec:appendix-nc-table}. 

% \begin{table}[ht]
%     \centering
%     \caption{Comparison of \( \overline{Nu} \) for different methods. }
%     \label{tab:nc_Nu}
%     % \vspace{0.5em}
%     \footnotesize
%     \begin{tabular}{ccccc}
%         \toprule
%         References (Grid)& \(10^6\) & \(10^7\) & \(5 \times 10^7\) & \(10^8\) \\
%         \midrule
%         CFD, Kalita et al.(81x81) \citep{kalita2001fully}& 8.829 & 16.52 & - & - \\
%         CFD, Chiu(75$\times$75) \citep{chiu2018CFD} & 8.785 & 16.46 & - & - \\
%         CFD, Chiu (202$\times$202)\citep{chiu2018CFD} & - & 16.50 & 25.27 & 30.34 \\
%         Ours (75$\times$75) & 8.695 & 16.31 & - & - \\
%         Ours (202$\times$202) & - & 16.34 & 24.87 & 29.50 \\
%         \bottomrule
%     \end{tabular}
% \end{table}

\begin{table}[ht]
\centering
\vspace{0.3em}
% \footnotesize
\small
\caption{Comparison of average Nusselt number ($\overline{Nu}$) and maximum vertical velocity ($v_{max}$) at various $Ra$ for different methods.}
\label{tab:numerical_comparison}
\setlength{\tabcolsep}{3pt}
\begin{tabular}{cccccc}
\toprule
References & \multirow{2.5}{*}{Parameters} & \multicolumn{4}{c}{$Ra$} \\
\cmidrule(lr){3-6}
(Mesh/Sample points) & & $10^6$ & $10^7$ & $5 \times 10^7$ & $10^8$ \\
\midrule
CFD, Kalita et al.  
& $\overline{Nu}$ & 8.829 & 16.52 & -- & -- \\
$\left(81 \times 81 \right)$\citep{kalita2001fully} & $v_{max}$ & 221.66 & 696.2 & -- & -- \\
\addlinespace[0.3em]
CFD, Chiu  
& $\overline{Nu}$ & 8.785 & 16.46 & -- & -- \\
$\left(75 \times 75 \right)$\citep{chiu2018CFD} & $v_{max}$ & 215.80 & 681.73 & -- & -- \\
\addlinespace[0.3em]
CFD, Chiu  
& $\overline{Nu}$ & -- & 16.50 & 25.27 & 30.34 \\
$\left(202 \times 202 \right)$ \citep{chiu2018CFD} & $v_{max}$ & -- & 695.22 & 1543.92 & 2202.26 \\
\addlinespace[0.3em]
Ours 
& $\overline{Nu}$ & 8.695 & 16.31 & -- & -- \\
$\left(75 \times 75 \right)$ & $v_{max}$ & 216.42 & 674.78 & -- & -- \\
\addlinespace[0.3em]
Ours 
& $\overline{Nu}$ & -- & 16.34 & 24.87 & 29.50 \\
$\left(202 \times 202 \right)$ & $v_{max}$ & -- & 696.33 & 1443.62 & 2129.92 \\
\bottomrule
\end{tabular}
\end{table}

% \begin{table}[ht]
%     \centering
%     \caption{Comparison of $v_{max}$ for different methods.}
%     \label{tab:nc_v}
%     % \vspace{0.5em}
%     \footnotesize
%     \begin{tabular}{ccccc}
%     \toprule
%         References (Grid) & \(10^6\) & \(10^7\) & \(5 \times10^7\)& \(10^8\) \\
%         \midrule
%         CFD, Kalita et al. (81$\times$81)\citep{kalita2001fully} & 221.66 & 696.2 &- & - \\
%         CFD, Chiu (75$\times$75)\citep{chiu2018CFD} & 215.801 & 681.730 & - & -\\
%         CFD, Chiu  (202$\times$202)\citep{chiu2018CFD} & - & 695.224 & 1543.921 & 2202.258\\
%         Ours (75$\times$75) & 216.423 & 674.779& - &-\\
%         Ours (202$\times$202) & - & 696.334 & 1443.615 &2129.916\\
%         \bottomrule
%     \end{tabular}
% \end{table}

Table \ref{tab:nc_error_table} presents a quantitative evaluation of the FFV-PINN model's accuracy in simulating natural convection across a range of $Ra$ ($Ra = 10^6$ to $10^8$) by utilizing CFD solutions as ground truth\citep{chiu2018CFD}. The model's performance is assessed using MSE and relative $L_2$ error for the $u$-velocity, $v$-velocity, and temperature ($T$) fields.  Training time and sample points are also provided. The FFV-PINN model exhibits excellent performance across the $Ra$ considered.  While both the MSE and relative $L_2$ error metrics generally increase with increasing $Ra$, the magnitude of these errors remains remarkably low, even at the highest $Ra$ value of $10^8$.  This indicates the model's ability to accurately capture the complex interplay of buoyancy and diffusion that characterizes natural convection, even as the flow transitions towards a more nonlinear regime. The training time remains consistently low across all $Ra$ ranges, varying between 150s and 231s, thus confirming the efficiency of the FFV-PINN model. Last, it is worth noting that, to the best of the authors' knowledge, this work represents the first successful application of a PINN-based method to natural convection problems at $Ra = 10^8$. This achievement underscores the capability of the FFV-PINN model to address highly complex flow regimes, marking a significant advancement in the application of machine learning techniques to challenging fluid dynamics problems.
\begin{table}[ht]
\centering
\caption{Quantitative evaluation of model accuracy for natural convection at various $Ra$}
\label{tab:nc_error_table}
% \vspace{0.5em}
\resizebox{\textwidth}{!}{
\begin{tabular}{ccccccccc}
\toprule
\multirow{2.5}{*}{$Ra$} & \multicolumn{3}{c}{MSE} & \multicolumn{3}{c}{Relative $L_2$ Error} & \multirow{2.5}{*}{Time (s)} & \multirow{2.5}{*}{Sample points} \\
\cmidrule(lr){2-4} \cmidrule(lr){5-7}
 & $u$ & $v$ & $T$ & $u$ & $v$ & $T$ & & \\
\midrule
$10^6$ & 1.42e-06 & 2.01e-06 & 6.25e-06 & 2.52e-02 & 2.09e-02 & 4.51e-03 & 164 & 50$\times$50 \\
$10^6$ & 1.12e-06 & 1.44e-06 & 4.95e-06 & 2.24e-02 & 1.76e-02 & 4.01e-03 & 154 & 75$\times$75 \\

$10^7$ & 4.30e-06 & 4.50e-06 & 6.34e-06 & 6.90e-02 & 4.05e-02 & 6.03e-03 & 150 & 75$\times$75 \\
$10^7$ &  3.47e-06 & 1.94e-06 & 9.84e-06 & 6.18e-02  & 2.62e-02 & 5.65e-03 & 225 & 202$\times$202 \\
$5\times10^7$ & 3.55e-06 & 3.45e-06& 1.38e-05 &8.69e-02 & 4.19e-02 & 6.67e-03 & 230 & 202$\times$202 \\
$10^8$ & 1.56e-05 & 1.09e-05 & 1.37e-05 & 2.12e-01 & 8.07e-02 & 6.65e-03 & 231 & 202$\times$202 \\
\bottomrule
\end{tabular}
}
\end{table}

% The table shows an increase in mesh with increasing $Ra$, reflecting the need for a finer resolution to capture the finer flow structures at higher $Ra$. The model effectively adapts to this requirement, maintaining accuracy even with increased mesh complexity.

\section{Conclusion}\label{sec:remarks}
In this study, we present the FFV-PINN, an innovative framework that synergistically integrates a simplified FVM with the residual correction loss term. This framework effectively mitigates two primary shortcomings commonly associated with vanilla PINNs: the inadequate enforcement of physical constraints in the loss function due to the omission of information from neighboring points, and the challenges related to poor convergence during the training process.

Traditional FVM is well-known for its inherent mechanisms that enforce conservation laws, which often leads to solutions that are not only stable but also reliable. However, selecting an appropriate discretization scheme for the convection term remains non-trivial, necessitating a comprehensive understanding of the underlying physical phenomena as well as the strengths and limitations of the available numerical methods. In FFV-PINN, we simplify this process by directly leveraging neural network outputs on control surfaces, strengthening the enforcement of physical constraints while reducing the complexity of discretization. Moreover, the residual correction loss plays a crucial role in improving both stability and convergence, making the training process more robust. By guiding the training direction of the neural network, this term ensures that updates are executed in a manner that aligns with the fulfillment of physical constraints. As a result, the FFV-PINN framework mitigates the issues of poor convergence and fosters a highly efficient training environment.

To validate the effectiveness of the proposed FFV-PINN framework, we conducted a comprehensive evaluation by applying it to a series of challenging benchmark problems, including 2D steady and unsteady lid-driven cavity flows, 3D steady lid-driven cavity flow, backward-facing step flow, and natural convection. The results, when compared with existing literature, demonstrate that FFV-PINN significantly enhances both accuracy and computational efficiency. This improvement is particularly evident in the framework's ability to achieve higher solution precision while requiring substantially less training time. Without incorporating additional data, previous studies on PINNs have only been able to solve the lid-driven cavity flow problem up to $Re$ = 5000, while requiring substantial computational resources and time. In contrast, our FFV-PINN successfully extends the solution to $Re$ = 10000, a highly non-linearity regime that has not been achieved by other PINN methods, while reducing training time significantly. This achievement underscores the capability of FFV-PINN to address complex fluid dynamics scenarios. Furthermore, FFV-PINN also demonstrates high accuracy and computational efficiency when solving 2D unsteady and 3D flow problems, underscoring the model’s capability and robustness.

Despite the promising results achieved with the FFV-PINN, it is important to acknowledge that the framework still has certain limitations. Specifically, for the lid-driven cavity flow at $Re$ = 10000, the FFV-PINN fails to effectively capture the expected third small-scale vortex structure in the lower-left corner of the cavity. This limitation underscores the necessity for further advances in capturing small-scale phenomena, which are critical for accurately representing turbulent flow dynamics. Future research will focus on addressing this challenge by exploring strategies such as assigning greater weight to small-scale solutions during training. Additionally, we will consider integrating multi-grid techniques to enable adaptive refinement of the model's grid or resolution in regions where small-scale phenomena are prominent. 

Beyond these directions, another promising avenue lies in extending the FFV-PINN framework to incorporate higher-order FVM. Such extensions would involve reformulating the flux at cell interfaces and the discretization of diffusion terms using a higher-order interpolation scheme. While this would introduce additional computational complexity, it is expected to improve solution accuracy and better capture fine-scale features, particularly in high $Re$ number. We consider this a valuable direction for future work, whereby one might opt for specific schemes to achieve the highest accuracy within acceptable computational resource limits, as is common in traditional scientific computing. % %% The Appendices part is started with the command \appendix;
% %% appendix sections are then done as normal sections
\section*{Acknowledgment}
Chang Wei and Yuchen Fan would like to acknowledge support from the China Scholarship Council for the scholarship to conduct research at Agency for Science, Technology and Research (A*STAR).

This research is partially supported by the National Research Foundation, Singapore, and Civil Aviation Authority Singapore under the Aviation Transformation Programme - "ATM-Met Integration of Convective Weather Forecast and Impact Forecast Solutions Supporting Singapore Air Traffic Operations [Award No. ATP2.0\_ATM-MET\_I2R], and the Singapore Ministry of Health’s National Medical Research Council under under its National Epidemic Preparedness and Response R\&D Funding Initiative (MOH-001041) Programme for Research in Epidemic Preparedness and REsponse (PREPARE) - "Development of Risk Assessment Model for Infectious Respiratory Diseases Considering Human Mobility [Award No. PREPARE-CS1-2022-004].
% under xxx: [Award No. xxx].

\clearpage
\appendix

\section{}\label{sec: appendix coefficient fvm}
% When employing the SOU scheme for the momentum equation, the coefficients and source term of the Eq. \ref{disc_eq} are as follows:
% \begin{subequations} 
% \small
% \begin{align}
%  &a_E = -\frac{1}{Re} - \frac{3}{2} max[-F_e,0] -  \frac{1}{2} max[-F_w,0]\\
%  &a_W = -\frac{1}{Re} - \frac{3}{2} max[F_w,0] -  \frac{1}{2} max[F_e,0]\\
%  &a_N = -\frac{1}{Re} - \frac{3}{2} max[-F_n,0] -  \frac{1}{2} max[-F_s,0]\\
%  &a_S = -\frac{1}{Re} - \frac{3}{2} max[F_s,0] -  \frac{1}{2} max[F_n,0]\\
%  &a_{EE} =  \frac{1}{2} max[-F_e,0]\\
%  &a_{WW} =   \frac{1}{2} max[F_w,0]\\
%  &a_{NN} =  \frac{1}{2} max[-F_n,0]\\
%  &a_{SS} =   \frac{1}{2} max[F_s,0]\\
%  &a_P = \frac{4}{Re} + \frac{3}{2} \left( max[F_e,0] + max[-F_w,0] +max[F_n,0] + max[-F_s,0] \right ) \\
%  % & b = \frac{-\left (p_E - p_W \right )\Delta y}{2} or \frac{-\left (p_N - p_S \right )\Delta x}{2}\\
%  & b = 
%  \begin{cases}
% \frac{\left (p_E - p_W \right )\Delta y}{2}& \text{if } \phi = u \\
% \frac{\left (p_N - p_S \right )\Delta x}{2}& \text{if } \phi = v 
% \end{cases}
% \end{align}
% \end{subequations}

% \clearpage
% \section{}\label{sec: appendix simplified FVM}
When employing the simplified FVM for the momentum equation, the coefficients and source term of Eq. \ref{eq_dis} are as follows:
\begin{subequations} \label{eq_coff}
\begin{align}
 &a_E = a_W = a_N = a_S = -\frac{1}{Re} \\
 &a_e = u_e\Delta y\\
 &a_w = -u_w\Delta y\\
 &a_n = v_n\Delta x\\
 &a_s = -v_s\Delta x\\
 &a_P = \frac{4}{Re} \\
 & b = 
 \begin{cases}
\frac{\left (p_E - p_W \right )\Delta y}{2}& \text{if } \phi = u \\
\frac{\left (p_N - p_S \right )\Delta x}{2}& \text{if } \phi = v 
\end{cases}
 \end{align}
\end{subequations}
% \clearpage

\section{}\label{sec: appendix analysis}
\begin{itemize}
    \item \textit{\textbf{CD Scheme}}
\end{itemize}

The spatial derivative $\frac{\partial \phi}{\partial x} $ can be expressed as follows:
\begin{equation}\label{partial_spatial_cd}
    \frac{\partial \phi}{\partial x} = \frac{\phi_E - \phi_W}{2h} = \frac{\phi (x + h) - \phi (x - h)}{2h}
\end{equation}

Subsequently, the Fourier transform is applied to Eq. \ref{partial_spatial_cd}.
\begin{equation}
    2ikh\hat{\phi}(k) =  e^{ikh} \hat{\phi}(k) - e^{-ikh} \hat{\phi}(k) 
\end{equation}

Rearranging the above equation yields:
\begin{equation}
    2ikh = e^{ikh}  - e^{-ikh} 
\end{equation}

By defining effective wavenumber $k^{'}  \cong  k$, and multiplying both sides of the above equation by $-i$, we obtain 
\begin{equation}\label{kh_cd}
    k^{'}h = \frac{-i}{2}\left( e^{ikh}  - e^{-ikh} \right)
\end{equation}

The real and imaginary parts of Eq. \ref{kh_cd} can then be expressed as:
\begin{subequations}
\begin{align}
& \Re \left( k^{'}h \right) = \Re \left[ \frac{-i}{2}\left( e^{ikh}  - e^{-ikh} \right)\right] = \sin\left(kh\right) \\
& \Im \left( k^{'}h \right) =\Im \left[ \frac{-i}{2}\left( e^{ikh}  - e^{-ikh} \right) \right] = 0
\end{align}
\end{subequations}

\begin{itemize}
    \item \textit{\textbf{FOU Scheme}}
\end{itemize}

The spatial derivative $\frac{\partial \phi}{\partial x} $ can be expressed as:
\begin{equation}\label{partial_spatial_fou}
    \frac{\partial \phi}{\partial x} = \frac{\phi_e - \phi_w}{h} = \frac{\phi (x) - \phi (x - h)}{h}
\end{equation}

Subsequently, the Fourier transform is applied to Eq. \ref{partial_spatial_fou}.
\begin{equation}
    ikh\hat{\phi}(k) =  \hat{\phi}(k) - e^{-ikh} \hat{\phi}(k) 
\end{equation}

Rearranging the above equation yields:
\begin{equation}
    ikh = 1  - e^{-ikh} 
\end{equation}

By defining effective wavenumber $k^{'}  \cong  k$, and multiplying both sides of the above equation by $-i$, we obtain 
\begin{equation}\label{kh_fou}
    k^{'}h = -i\left( 1 - e^{-ikh} \right)
\end{equation}

The real and imaginary parts of Eq. \ref{kh_fou} can then be expressed as:
\begin{subequations}
\begin{align}
& \Re \left( k^{'}h \right) = \Re \left[ -i\left( 1 - e^{-ikh} \right)\right] = \sin(kh) \\
& \Im \left( k^{'}h \right) =\Im \left[ -i\left( 1 - e^{-ikh} \right) \right] = \cos(kh) - 1
\end{align}
\end{subequations}

\begin{itemize}
    \item \textit{\textbf{SOU Scheme}}
\end{itemize}

The spatial derivative $\frac{\partial \phi}{\partial x} $ can be expressed as follows:
\begin{equation}\label{partial_spatial_sou}
    \frac{\partial \phi}{\partial x} = \frac{\phi_e - \phi_w}{h} = \frac{3\phi (x) - 4\phi (x - h) + \phi (x - 2h)}{2h}
\end{equation}

Subsequently, the Fourier transform is applied to Eq. \ref{partial_spatial_sou}.
\begin{equation}
    2ikh\hat{\phi}(k) =  \left(3 - 4e^{-ikh} + e^{-i2kh}\right) \hat{\phi}(k)
\end{equation}

Rearranging the above equation yields:
\begin{equation}
    2ikh =  \left(3 - 4e^{-ikh} + e^{-i2kh}\right) 
\end{equation}

By defining effective wavenumber $k^{'}  \cong  k$, and multiplying both sides of the above equation by $-i$, we obtain 
\begin{equation}\label{kh_sou}
    k^{'}h = \frac{-i}{2}\left(3 - 4e^{-ikh} + e^{-i2kh}\right) 
\end{equation}

The real and imaginary parts of Eq. \ref{kh_sou} can then be expressed as:
\begin{subequations}
\begin{align}
& \Re \left( k^{'}h \right) = \frac{1}{2} \left( 4\sin(kh) - \sin(2kh) \right) \\
& \Im \left( k^{'}h \right) =  \frac{-1}{2} \left( 3 - 4\cos(kh) + \cos(2kh) \right)
\end{align}
\end{subequations}

\section{}\label{sec:appendix-nc-table}
\begin{sideways}  
  \begin{minipage}{\textheight} 
    \centering
    % \big
    \captionof{table}{Comparison of numerical results at different Rayleigh numbers ($Ra$). 
             Shown are the maximum horizontal velocity ($u_{\max}$) with its location, 
             maximum vertical velocity ($v_{\max}$) with its location, 
             and average Nusselt number ($\overline{Nu}$).}
    % \vspace{0.5em}
    \renewcommand{\arraystretch}{2}      
    \setlength{\tabcolsep}{4pt}            
    \resizebox{\textwidth}{!}{%
      \begin{tabular}{cccccccccccccc} %
        \toprule
        \multirow{2}{*}{Method (Mesh/Sample points)} & \multicolumn{3}{c}{$Ra=10^6$} & \multicolumn{3}{c}{$Ra=10^7$} & \multicolumn{3}{c}{$Ra=5\!\times\!10^7$} & \multicolumn{3}{c}{$Ra=10^8$} \\
        \cmidrule(lr){2-4} \cmidrule(lr){5-7} \cmidrule(lr){8-10} \cmidrule(lr){11-13}
         & $u_{max}$ & $v_{max}$ & $\overline{Nu}$ & $u_{max}$ & $v_{max}$ & $\overline{Nu}$ & $u_{max}$ & $v_{max}$ & $\overline{Nu}$ & $u_{max}$ & $v_{max}$ & $\overline{Nu}$ \\
        \midrule
        CFD, Kalita et al.(81$\times$81)\citep{kalita2001fully}
            & 65.332   & 221.658   & \multirow{2}{*}{8.829} 
            & 155.82   & 696.238   & \multirow{2}{*}{16.517}
            & -        & -         & \multirow{2}{*}{-}     
            & -        & -         & \multirow{2}{*}{-}
             \\
            & (0.850)  & (0.038)   & {} 
            & (0.863)  & (0.025)   & {} 
            & -        & -         & {}     
            & -        & -         & {} \\
        \multirow{2}{*}{CFD, Chiu (75$\times$75)\citep{chiu2018CFD}} 
            & 64.883   & 215.801   & \multirow{2}{*}{8.785} 
            & 148.4308 & 681.7296  & \multirow{2}{*}{16.456} 
           & - & -  & \multirow{2}{*}{-} 
            & - & -  & \multirow{2}{*}{-}     \\
         & (0.847)  & (0.034)   & {} 
            & (0.873)  & (0.020)   & {} 
            & -  &  -  & {} 
            & -  &  -  & {}
                \\

        \multirow{2}{*}{CFD, Chiu (202$\times$202)\citep{chiu2018CFD}}
            & -  & -  & \multirow{2}{*}{-} 
            & 148.51 & 695.224  & \multirow{2}{*}{16.501} 
           & 287.326  & 1543.921  & \multirow{2}{*}{25.277} 
            & 333.533  & 2202.258  & \multirow{2}{*}{30.341}     \\
         & -  & -  & {} 
            & (0.882)  & (0.025)   & {} 
            & (0.938)  & (0.012)   & {} 
            & (0.932)  & (0.012)   & {}
                \\
        \multirow{2}{*}{Ours (75$\times$75)} 
            & 64.885   & 216.423   & \multirow{2}{*}{8.695} 
            & 133.506  & 674.779   & \multirow{2}{*}{16.312}
            & - & -  & \multirow{2}{*}{-} 
            & - & - & \multirow{2}{*}{-}
              \\
         & (0.848)  & (0.040)   & {} 
            & (0.889)  & (0.020)   & {}
            & -  & -   & {} 
            & -  & -  & {}
              \\
        \multirow{2}{*}{Ours (202$\times$202)} 
            & -   & -  & \multirow{2}{*}{-}
            & 134.611 & 696.334  & \multirow{2}{*}{16.342}
            & 257.201 & 1443.615  & \multirow{2}{*}{24.872} 
            & 303.094 & 2129.916 & \multirow{2}{*}{29.504}
              \\
         & -  & -  & {} 
            & (0.879)  & (0.020)   & {}
            & (0.939)  & (0.010)   & {} 
            & (0.919)  & (0.010)   & {}
              \\
        \bottomrule
    \end{tabular}%
    }
  \end{minipage}
\end{sideways}


\bibliographystyle{elsarticle-harv} 

\bibliography{cas-refs}





% \end{thebibliography}
\end{document}